\documentclass[journal]{IEEEtran}
\normalsize

\ifCLASSINFOpdf
\else
\fi

\usepackage{graphicx,cite,epsfig,amssymb,amsmath,multirow}

\usepackage{array}

\usepackage{mdwmath}
\usepackage{mdwtab}

\usepackage{eqparbox}

\usepackage{stfloats}

\usepackage{setspace}

\begin{document}
%
\title{
{Novel 3D Geometry-Based Stochastic Models for Non-Isotropic MIMO Vehicle-to-Vehicle Channels}
}
%
%

\author{
        Yi~Yuan,~\IEEEmembership{}
        Cheng-Xiang~Wang,~\IEEEmembership{Senior Member, IEEE,} 
        Xiang~Cheng,~\IEEEmembership{Senior Member, IEEE,} \\
        Bo~Ai,~\IEEEmembership{Senior Member, IEEE,}
        and~David~I.~Laurenson,~\IEEEmembership{Member, IEEE}  

\thanks{Manuscript received March 10, 2013; revised August 2, 2013; accepted October 16, 2013. The associate editor coordinating the review of this paper and approving it for publication was Md Jahangir Hossain.}
\thanks{This paper was presented in part at IEEE VTC'10-Fall, Ottawa, Canada, Sep. 2010.}
\thanks{Y.~Yuan and C.-X.~Wang (corresponding author) are with the Joint Research Institute for Signal and Image Processing, School of Engineering \& Physical Sciences, Heriot-Watt University, Edinburgh, EH14 4AS, U.K. (e-mail: \{yy120, cheng-xiang.wang\}@hw.ac.uk).}
\thanks{X.~Cheng is with the Research Institute for Modern Communications, School of Electronics Engineering \& Computer Science, Peking University, Beijing, 100871, China (e-mail: xiangcheng@pku.edu.cn).}
\thanks{B.~Ai is with the State Key Laboratory of Rail Traffic Control and Safety, Beijing Jiaotong University, Beijing, 100044, China (e-mail: boai@bjtu.edu.cn).}
\thanks{D.~I.~Laurenson is with the Joint Research Institute for Signal and Image Processing, Institute for Digital Communications, University of Edinburgh, Edinburgh, EH9 3JL, U.K. (e-mail: Dave.Laurenson@ed.ac.uk).}
\thanks{This work was supported by the Opening Project of the Key Laboratory of Cognitive Radio and Information Processing (Guilin University of Electronic Technology), Ministry of Education (No.: 2013KF01), the National Natural Science Foundation of China (Grant~No.~61222105 \& 61101079), Beijing Municipal Natural Science Foundation (Grant~No.~4112048), and the Science Foundation for the Youth Scholar of Ministry of Education of China (Grant~No. 20110001120129).}
}

%
%

\markboth{IEEE Transactions on Wireless Communications}%
{Yi Yuan \MakeLowercase{\textit{et al.}}: Novel 3D Geometry-Based Stochastic Models for Non-Isotropic MIMO Vehicle-to-Vehicle Channels}

%



\maketitle

\begin{abstract}
This paper proposes a novel three-dimensional (3D) theoretical regular-shaped geometry-based stochastic model (RS-GBSM) and the corresponding sum-of-sinusoids (SoS) simulation model for non-isotropic multiple-input multiple-output (MIMO) vehicle-to-vehicle (V2V) Ricean fading channels. The proposed RS-GBSM, combining line-of-sight (LoS) components, a two-sphere model, and an elliptic-cylinder model, has the ability to study the impact of the vehicular traffic density (VTD) on channel statistics, and jointly considers the azimuth and elevation angles by using the von Mises Fisher distribution. Moreover, a novel parameter computation method is proposed for jointly calculating the azimuth and elevation angles in the SoS channel simulator. 
Based on the proposed 3D theoretical RS-GBSM and its SoS simulation model, statistical properties are derived and thoroughly investigated. 
The impact of the elevation angle in the 3D model on key statistical properties is investigated by comparing with those of the corresponding two-dimensional (2D) model. It is demonstrated that the 3D model is more accurate to characterize real V2V channels, in particular for pico cell scenarios. Finally, close agreement is achieved between the theoretical model, SoS simulation model, and simulation results, demonstrating the utility of the proposed models. 
\end{abstract}


%

\IEEEpeerreviewmaketitle

\section{Introduction}

\IEEEPARstart{I}{n} recent years, vehicle-to-vehicle (V2V) communications \cite{IEEE80211p} have been encountered in many new applications, such as wireless mobile ad hoc peer-to-peer networks \cite{Chen05, Hartenstein08} cooperative systems \cite{Wang10}, \cite{Patzold11}, and intelligent transportation systems. In V2V communication systems, both the transmitter (Tx) and receiver (Rx) are in motion and equipped with low elevation antennas. This is different from conventional fixed-to-mobile (F2M) cellular radio systems, where only one terminal moves. Moreover, multiple-input multiple-output (MIMO) technologies, where multiple antennas are deployed at both the Tx and Rx \cite{Almers07}, have widely been adopted in advanced F2M cellular systems and have also been receiving more and more attention in V2V systems \cite{Bakhshi12}.

In order to evaluate the performance of a V2V communication system, accurate channel models are indispensable. Existing channel models for F2M communications systems cannot be used directly for the design of V2V systems. 
V2V channel models available in the literature \cite{Maurer08, Acosta07, XCheng09WCMC, Akki86, Akki94, Zajic08TVT, XCheng09TWC, XCheng10VTC, Zajic083d, Zajic093d, Samarasinghe2010, Wu10, Cheng12, Wang09, Boeglen2011, Karedal09TWC} can be classified as geometry-based deterministic models (GBDMs) \cite{Maurer08} and stochastic models, which can further be categorized as non-geometry-based stochastic models (NGSMs) \cite{Acosta07} and geometry-based stochastic models (GBSMs) \cite{XCheng09WCMC, Akki86, Akki94, Zajic08TVT, XCheng09TWC, XCheng10VTC, Zajic083d, Zajic093d, Samarasinghe2010, Wu10, Cheng12, Wang09, Boeglen2011, Karedal09TWC}. Furthermore, GBSMs can be classified as regular-shaped GBSMs (RS-GBSMs) \cite{Akki86, Akki94, Zajic08TVT, XCheng09TWC, XCheng10VTC, Zajic083d, Zajic093d, Samarasinghe2010, Wu10, Cheng12} and irregular-shaped GBSMs (IS-GBSMs) \cite{Wang09, Boeglen2011, Karedal09TWC, Boban11JSAC}, depending on whether effective scatterers are located on regular shapes, e.g., one-ring, two-ring, ellipses, or irregular shapes. 


RS-GBSMs \cite{Akki86, Akki94, Zajic08TVT, XCheng09TWC, XCheng10VTC, Zajic083d, Zajic093d, Samarasinghe2010, Wu10, Cheng12} have widely been used to mimic V2V channels due to their convenience for theoretical analysis of channel statistics. To preserve the mathematical tractability, RS-GBSMs assume that all the effective scatterers are located on regular shapes. Akki and Haber were the first to propose a two-dimensional (2D) RS-GBSM \cite{Akki86} and investigate corresponding statistical properties for narrowband isotropic scattering single-input single-output (SISO) V2V Rayleigh channels \cite{Akki94}. 
In \cite{Zajic08TVT}, the authors proposed a 2D two-ring RS-GBSM with both single- and double-bounced rays for narrowband non-isotropic scattering MIMO V2V Ricean channels. In \cite{XCheng09TWC}, the authors proposed an adaptive RS-GBSM consisting of two rings and one ellipse also with both single- and double-bounced rays for narrowband non-isotropic MIMO V2V Ricean channels. As 2D models assume that waves travel only in the horizontal plan, they neglect signal variations in the vertical plane and are valid only when the Tx and Rx are sufficiently separated. In reality, waves do travel in three dimensions. 
Therefore, a three-dimensional (3D) two-cylinder RS-GBSM was developed for narrowband non-isotropic scattering MIMO V2V channels in \cite{Zajic083d}. It was further extended to a wideband one in \cite{Zajic093d}. Other 3D V2V channel models include a 3D two-sphere RS-GBSM for narrowband non-isotropic SISO V2V channels \cite{Samarasinghe2010} and a 3D two-concentric-quasi-sphere RS-GBSM for wideband non-isotropic MIMO V2V channels \cite{Wu10}.

The aforementioned 3D RS-GBSMs \cite{Zajic083d, Zajic093d, Samarasinghe2010, Wu10} all assumed that the azimuth angle and elevation angle are completely independent and thus analyzed them separately. Moreover, although the measurement campaigns in \cite{Acosta07} demonstrated that the vehicular traffic density (VTD) significantly affects the V2V channel statistical properties, the impact of the VTD on channel statistics was not considered in the existing 3D RS-GBSMs \cite{Zajic083d, Zajic093d, Samarasinghe2010, Wu10}. 

To fill the above research gaps, the first part of this paper proposes a novel theoretical 3D RS-GBSM, which is the combination of line-of-sight (LoS) components, a two-sphere model, and an elliptic-cylinder model \cite{XCheng10VTC}, for non-isotropic MIMO V2V channels. The proposed 3D RS-GBSM is sufficiently generic and adaptive to model various V2V channels in different scenarios. It is the first 3D RS-GBSM that has the ability to study the impact of the VTD on channel statistics, and jointly considers the azimuth and elevation angles by applying the von Mises-Fisher (VMF) distribution as the scatterer distribution. As the 3D theoretical RS-GBSM assumes infinite numbers of effective scatterers, which results in the infinite complexity, it cannot be implemented in practice. However, a theoretical model can be used as a starting point to design a realizable simulation model that considers limited numbers of scatterers and has a reasonable complexity. Hence, the second part of this paper concentrates on developing a corresponding 3D MIMO V2V sum-of-sinusoids (SoS) based simulation model with a novel parameter computation method. Note that the proposed models have already considered the effect of diffuse scattering \cite{Karedal09TWC} by using double-bounced rays. Also, the impact of vehicles as obstacles on the LoS obstruction, as studied in measurements \cite{Boban11JSAC} and \cite{Abbas13}, can be captured in our models by adjusting relevant model parameters, e.g., the Ricean factor.

Overall, the major contributions and novelties of this paper are summarized as follows:\begin{enumerate}
\item Based on the novel 3D theoretical RS-GBSM, comprehensive statistical properties are derived and thoroughly investigated, i.e., amplitude and phase probability density functions (PDFs), space-time (ST) correlation function (CF), Doppler power spectral density (PSD), envelope level crossing rate (LCR), and average fade duration (AFD). Meanwhile, some inaccurate expressions in \cite{XCheng10VTC} are corrected.
\item The impacts of the VTD and elevation angle on aforementioned channel statistical properties are investigated by comparing with those of the corresponding 2D model.
\item The corresponding SoS simulation model is proposed by considering a finite number of scatterers at the Tx and Rx. 
\item A novel parameter computation method, namely the method of equal volume (MEV), is proposed to calculate the azimuth and elevation angles of proposed SoS simulation model. It is the first method for 3D MIMO channel models jointly computing the azimuth and elevation angles. 
\item The statistical properties of our SoS simulation model are verified by comparing with those of the reference model and simulated results. The results show that the simulation model is an excellent approximation of the reference model according to their statistical properties. 
\end{enumerate}

The remainder of this paper is structured as follows. Section~II introduces a novel 3D theoretical RS-GBSM for non-isotropic narrowband MIMO V2V Ricean channels. In Section~III, the corresponding 3D simulation model is developed with parameters calculated by the MEV. Simulation results and analysis are unveiled in Section~IV. Finally, we draw conclusions in Section V.

\begin{figure} [t]%
\centering
\parbox{8.5cm}
{%
\centering \hspace{-0.307cm} \includegraphics[scale=0.38]{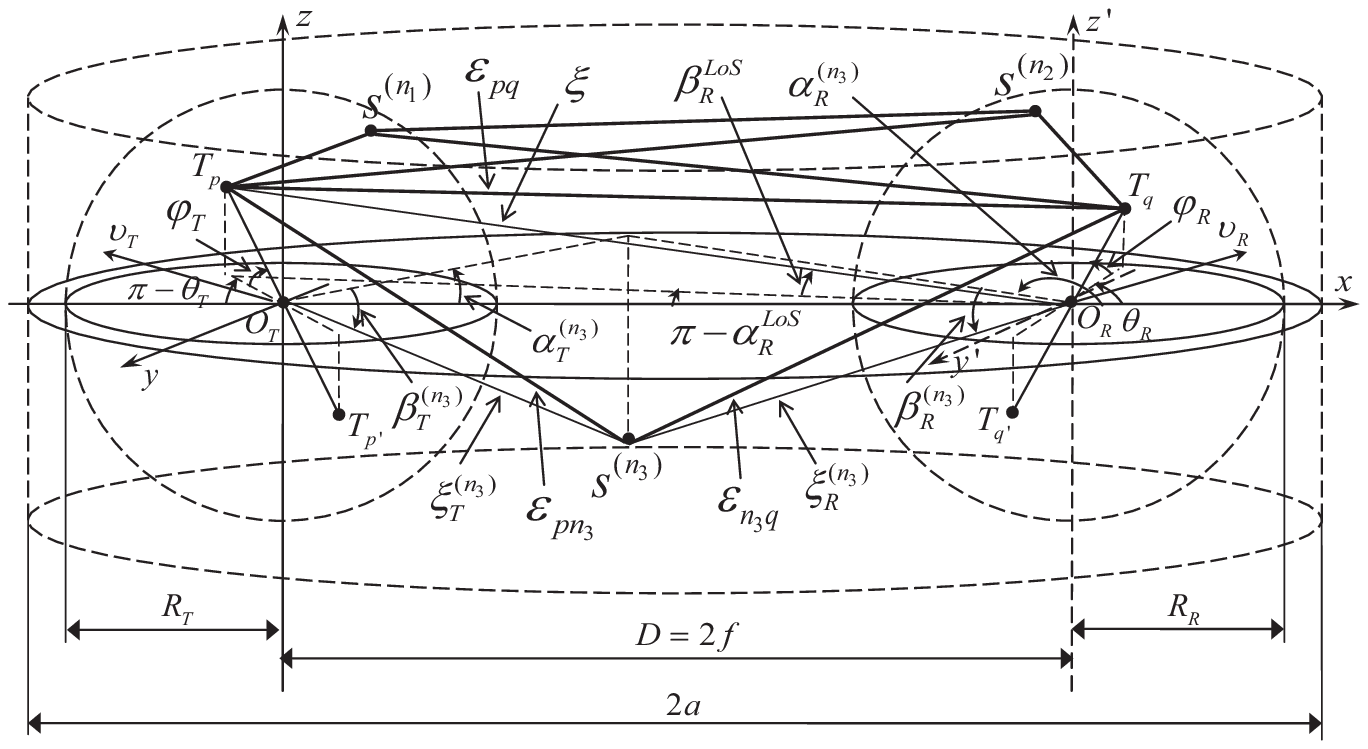} 
\vspace*{-0.8cm} 
\begin{center}
\caption{The proposed 3D MIMO V2V RS-GBSM combining a two-sphere model and an elliptic-cylinder model (only showing the detailed geometry of LoS components and single-bounced rays in the elliptic-cylinder model).}%
\end{center}
}%
\end{figure}

\begin{figure} [t]%
\centering
\parbox{8.5cm}
{%
\centering \hspace{-0.28cm} \includegraphics[scale=0.395]{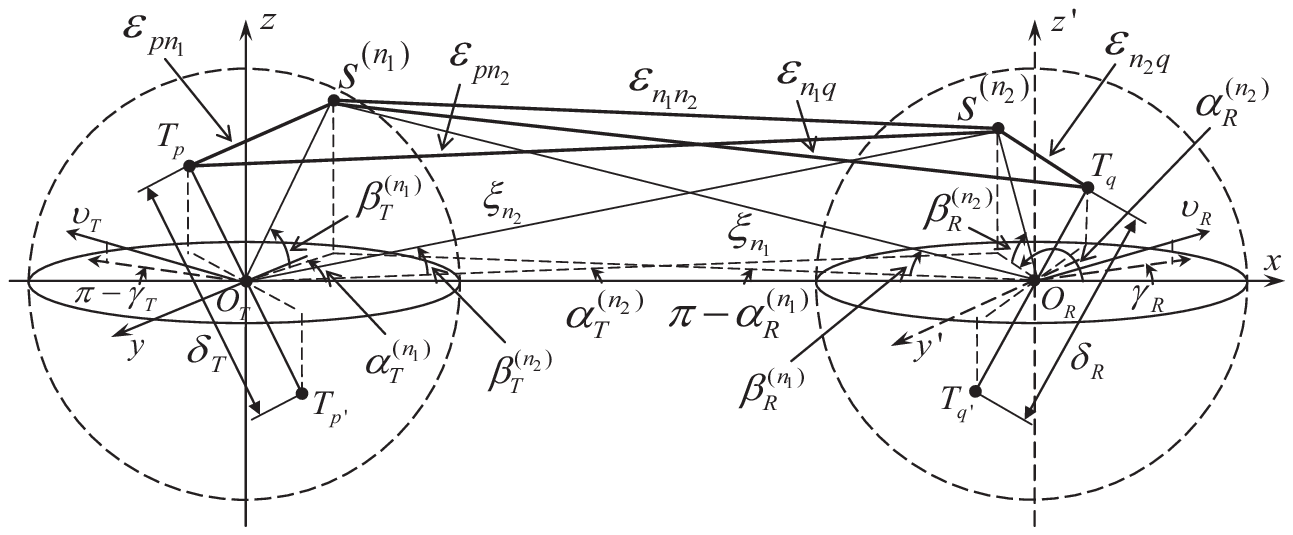} 
\vspace*{-0.8cm}
\begin{center}
\caption{The detailed geometry of the single- and double-bounced rays in the two-sphere model of the proposed 3D RS-GBSM.}%
\end{center}
}%
\end{figure}%

\begin{table*} [t]
\caption{Definition of parameters in Figs.~1 and 2.}
\centering
\renewcommand\arraystretch{1.0}
\begin{tabular} {|c|l|} 
\hline \small $D$ & \small distance between the centers of the Tx and Rx spheres \\ 
\hline \small $R_T$, $R_R$ & \small radius of the Tx and Rx spheres, respectively \\ 
\hline \small $a$, $f$ & \small semi-major axis and half spacing between two foci of the elliptic-cylinder, respectively \\
\hline \small $\delta_T$, $\delta_R$ & \small antenna element spacing at the Tx and Rx, respectively \\
\hline \small $\theta_T$, $\theta_R$ & \small orientation of the Tx and Rx antenna array in the $x$-$y$ plane, respectively \\
\hline \small $\varphi_T$, $\varphi_R$ & \small elevation of the Tx and Rx antenna array relative to the $x$-$y$ plane, respectively \\
\hline \small $\upsilon_T$, $\upsilon_R$ & \small velocities of the Tx and Rx, respectively. \\
\hline \small $\gamma_T$, $\gamma_R$ & \small moving directions of the Tx and Rx in the $x$-$y$ plane, respectively \\
\hline \small $\alpha_T^{(n_i)}$ & \small azimuth angle of departure (AAoD) of the waves that impinge on the effective \\
\small ($i=1,2,3$) & \small scatterers $s^{(n_i)}$ \\
\hline \small $\alpha_R^{(n_i)}$ & \small azimuth angle of arrival (AAoA) of the waves traveling from the effective \\
\small ($i=1,2,3$) & \small scatterers $s^{(n_i)}$ \\
\hline \small $\beta_T^{(n_i)}$ & \small elevation angle of departure (EAoD) of the waves that impinge on the effective \\
\small ($i=1,2,3$) & \small scatterers $s^{(n_i)}$ \\
\hline \scriptsize $\beta_R^{(n_i)}$ & \small elevation angle of arrival (EAoA) of the waves traveling from the effective \\
\small ($i=1,2,3$) & \small scatterers $s^{(n_i)}$ \\
\hline \small $\alpha_R^{LoS}$, $\beta_R^{LoS}$ & \small AAoA and EAoA of the LoS paths, respectively \\
\hline \small $\varepsilon_{pq}$, $\varepsilon_{pn_i}$, $\varepsilon_{n_1n_2}$, &  \\
\small $\varepsilon_{n_iq}$, $\xi$, $\xi_{T(R)}^{n_3}$, & \small distances $d\left(T_p,T_q\right)$, $d\left(T_p,s^{(n_i)}\right)$, $d\left(s^{(n_1)}, s^{(n_2)}\right)$, $d\left(s^{(n_i)}, T_q\right)$, $d\left(T_p, O_R\right)$,\\
\small $\xi_{n_1}$, $\xi_{n_2}$ & \small $d\left(O_T(O_R), s^{(n_3)}\right)$, $d\left(s^{(n_1)}, O_R\right)$, $d\left(O_T, s^{(n_2)}\right)$, respectively \\
\small ($i=1,2,3$) &  \\
\hline
\end{tabular}
\vspace{0.8cm} 
\end{table*}

\section{A Novel 3D MIMO V2V Theoretical RS-GBSM}
\subsection{Description of the 3D MIMO V2V theoretical RS-GBSM}
Let us consider a narrowband MIMO V2V communication system with $M_T$ transmit and $M_R$ receive omnidirectional antenna elements. The radio propagation environment is characterized by 3D effective scattering with LoS and non-LoS (NLoS) components between the Tx and Rx. Different from physical scatterers, an effective scatterer may include several physical scatterers which are unresolvable in delay and angle domains. 
Figs.~1 and~2 illustrate the proposed 3D RS-GBSM, which is the combination of LoS components, a single- and double-bounced two-sphere model, and a single-bounced elliptic-cylinder model. 
To consider the impact of the VTD on channel statistics, we need to distinguish between the moving vehicles around the Tx and Rx and the stationary roadside environments (e.g., buildings, trees, parked cars, etc.). Therefore, we use a two-sphere model to mimic the moving vehicles and an elliptic-cylinder model to depict the stationary roadside environments. It is worth mentioning that in order to significantly reduce the complexity of the 3D theoretical RS-GBSM, only the double-bounced rays via scatterers on the two-sphere model are considered because other double-bounced rays (via one scatterer on a sphere and the other one on the elliptic-cylinder) show similar channel statistics \cite{XCheng09TWC}. For readability purposes, Fig.~1 only shows the geometry of LoS components, and the single-bounced elliptic-cylinder model. The detailed geometry of the single- and double-bounced two-sphere model is given in Fig.~2. Note that in both Figs.~1 and~2, we adopted uniform linear antenna arrays with $M_T=M_R=2$ as an example. The proposed RS-GBSM can be extended with arbitrary numbers of antenna elements. By modeling effective scatterers, we assume that the two-sphere model defines two spheres of effective scatterers, one around the Tx and the other around the Rx. Suppose there are $N_{1}$ effective scatterers around the Tx lying on a sphere of radius $R_T$ and the $n_{1}$th ($n_{1}=1, ..., N_{1}$) effective scatterer is denoted by $s^{(n_{1})}$. Similarly, assume there are $N_{2}$ effective scatterers around the Rx lying on a sphere of radius $R_R$ and the $n_{2}$th ($n_{2}=1, ..., N_{2}$) effective scatterer is denoted by $s^{(n_{2})}$. For the elliptic-cylinder model, $N_{3}$ effective scatterers lie on an elliptic-cylinder with the Tx and Rx located at the foci and the $n_{3}$th ($n_{3}=1, ..., N_{3}$) effective scatterer is denoted by $s^{(n_{3})}$. The parameters in Figs.~1 and~2 are defined in Table~I. Note that the reasonable assumptions $D\gg{\mbox{max}}\{R_T, R_R\}$ and ${\mbox{min}}\{R_T, R_R, a-f\}\gg{\mbox{max}}\{\delta_T, \delta_R\}$ are applied in this theoretical model \cite{XCheng09TWC}.

\vspace{-0.0cm}
The 3D MIMO V2V channel is described by an $M_T \times M_R$ matrix of complex fading envelopes, i.e., $\mathbf{H} \left(t\right) = \left[ h_{pq} \left(t\right) \right]_{M_T \times M_R}$. The subscripts $p$ and $q$ denote the MIMO antenna elements. Therefore, the received complex fading envelope between the $p$th ($p=1,...,M_T$) Tx and the $q$th ($q=1,...,M_R$) Rx at the carrier frequency $f_c$ is a superposition of the LoS, single- and double-bounced components, and can be expressed as
\begin{align}
\label{eq:h_pq(t)} 
h_{pq}\left(t\right) &= h_{pq}^{LoS}\left(t\right) + \sum_{i=1}^{I}h_{pq}^{SB_i}\left(t\right) + h_{pq}^{DB}\left(t\right) 
\end{align}
where
\begin{subequations}
\label{eq:h_pq_L_S_D(t)}
\begin{align} 
h_{pq}^{LoS}(t) &= \sqrt{\frac{K}{K+1}}e^{-j2\pi f_c \tau_{pq}} \nonumber \\
&\times e^{j2\pi f_{T_{max}} t \cos\left(\alpha_T^{LoS}-\gamma_T\right)\cos{\beta_T^{LoS}}} \\ 
&\times e^{j2\pi f_{R_{max}} t \cos\left(\alpha_R^{LoS}-\gamma_R\right)\cos{\beta_R^{LoS}}} \nonumber
\end{align}
\begin{align} 
h_{pq}^{SB_i}\left(t\right) & = \sqrt{\frac{\eta_{SB_{i}}}{K+1}}\lim_{N_{i}\to\infty}\sum_{n_{i}=1}^{N_{i}}\frac{1}{\sqrt{N_{i}}}e^{j\left(\psi_{n_{i}}-2\pi f_c \tau_{pq,n_{i}}\right)} \nonumber \\
&\times e^{j2\pi f_{T_{max}} t \cos\left(\alpha_T^{(n_{i})}-\gamma_T \right)\cos{\beta_T^{(n_i)}}} \\
&\times e^{j2\pi f_{R_{max}} t \cos \left(\alpha_R^{(n_{i})}-\gamma_R \right)\cos{\beta_R^{(n_i)}}} \nonumber
\end{align}
\begin{align} \nonumber 
h_{pq}^{DB}\left(t\right)&=\sqrt{\frac{\eta_{DB}}{K+1}} \\ \nonumber
&\times \lim_{N_{1},N_{2}\to\infty}\sum_{n_{1},n_{2}=1}^{N_{1},N_{2}} \frac{1}{\sqrt{N_{1}N_{2}}} e^{j\left(\psi_{n_{1},n_{2}}-2\pi f_c \tau_{pq,n_{1},n_{2}}\right)} \\ 
&\times e^{j2\pi f_{T_{max}} t \cos\left(\alpha_T^{(n_{1})}-\gamma_T\right)\cos{\beta_T^{(n_1)}}} \\ \nonumber
&\times e^{j2\pi f_{R_{max}} t \cos\left(\alpha_R^{(n_{2})}-\gamma_R\right)\cos{\beta_R^{(n_2)}}} 
\end{align}
\end{subequations}
with $\alpha_T^{LoS}\approx\beta_T^{LoS}\approx\beta_R^{LoS}\approx 0$, $\alpha_R^{LoS}\approx\pi$, $\tau_{pq}=\varepsilon_{pq}/c$, $\tau_{pq,n_{i}}=(\varepsilon_{pn_{i}}+\varepsilon_{n_{i}q})/c$, and $\tau_{pq,n_{1},n_{2}} = (\varepsilon_{pn_{1}} + \varepsilon_{n_{1}n_{2}} + \varepsilon_{n_{2}q})/c$. Here, $c$ is the speed of light, $K$ designates the Ricean factor, and $I=3$ which means there are three subcomponents for single-bounced rays, i.e., $SB_1$ from the Tx sphere, $SB_2$ from the Rx sphere, and $SB_3$ from the elliptic-cylinder. Power-related parameters $\eta_{SB_{i}}$ and $\eta_{DB}$ specify the amount of powers that the single- and double-bounced rays contribute to the total scattered power $1/(K+1)$. Note that these power-related parameters satisfy $\sum_{i=1}^{I} \eta_{SB_{i}} + \eta_{DB}=1$. The phases $\psi_{n_{i}}$ and $\psi_{n_{1},n_{2}}$ are independent and identically distributed (i.i.d.) random variables with uniform distributions over $\left[-\pi, \pi\right)$, $f_{T_{max}}$ and $f_{R_{max}}$ are the maximum Doppler frequencies with respect to the Tx and Rx, respectively. 
Note that we have corrected inaccurate expressions (2a) and (2c) in \cite{XCheng10VTC}, corresponding to (2a) and (2c) in this paper, respectively. 

Based on the law of cosines in appropriate triangles and small angle approximations (i.e., $\sin x \approx x$ and $\cos x \approx 1$ for small $x$), we have
\begin{subequations}
\begin{align} 
\varepsilon_{pq} &\approx \xi-\frac{\delta_R}{2\xi}\left[\frac{\delta_T}{2}\sin\varphi_T\sin\varphi_R-Q\cos\varphi_R\cos\theta_R \right]
\end{align}
\begin{align} 
\varepsilon_{pn_{1}} &\approx R_T-\frac{\delta_T}{2}\left[\sin\beta_T^{(n_1)}\sin\varphi_T \right.\\ \nonumber
&+\left. \cos\beta_T^{(n_1)} \cos\varphi_T\cos(\theta_T-\alpha_T^{(n_1)})\right]
\end{align} 
\begin{align} 
\varepsilon_{n_{1}q} &\approx \xi_{n_1}-\frac{\delta_R}{2\xi_{n_1}}\left[R_T\sin\beta_T^{(n_1)}\sin\varphi_R \right.\\ \nonumber
&- \left. Q_{n_1}\cos\varphi_R\cos(\alpha_R^{(n_1)}-\theta_R)\right]
\end{align} 
\begin{align} 
\varepsilon_{pn_{2}} &\approx \xi_{n_2}-\frac{\delta_T}{2\xi_{n_2}}\left[R_R\sin\beta_R^{(n_2)}\sin\varphi_T \right.\\ \nonumber
&+ \left. Q_{n_2}\cos\varphi_T\cos(\alpha_T^{(n_2)}-\theta_T)\right]
\end{align} 
\begin{align} 
\varepsilon_{n_{2}q} &\approx R_R-\frac{\delta_R}{2}\left[\sin\beta_R^{(n_2)}\sin\varphi_R \right.\\ \nonumber
&+ \left. \cos\beta_R^{(n_2)} \cos\varphi_R\cos(\theta_R-\alpha_R^{(n_2)})\right]
\end{align} 
\begin{align} \nonumber 
\varepsilon_{n_{1}n_{2}} &\approx \left\{\left[D-R_T\cos\alpha_T^{(n_1)}-R_R \cos(\alpha_R^{(n_{1})}-\alpha_R^{(n_{2})})\right]^2 \right.\\
&+ \left. \left[R_T\cos\beta_T^{(n_1)}-R_R\cos\beta_R^{(n_2)}\right]^2 \right\}^{1/2}
\end{align} 
\begin{align} \nonumber 
\varepsilon_{pn_{3}} &\approx \xi_T^{(n_3)}-\frac{\delta_T}{2\xi_T^{(n_3)}} \left[\xi_R^{(n_3)}\sin\beta_R^{(n_3)}\sin\varphi_T \right.\\
&+ \left. Q_{n_3}\cos\varphi_T\cos(\alpha_T^{(n_3)}-\theta_T)\right]
\end{align} 
\begin{align} \nonumber 
\varepsilon_{n_{3}q} &\approx \xi_R^{(n_3)}-\delta_R \left[\sin\beta_R^{(n_3)}\sin\varphi_R \right.\\
&+ \left. \cos\beta_R^{(n_3)}\cos\varphi_R\cos(\alpha_R^{(n_3)}-\theta_R)\right]
\end{align} 
\end{subequations}
where $\xi \approx Q \approx D-\frac{\delta_T}{2}\cos\varphi_T\cos\theta_T$, $\xi_{n_1}=\sqrt{Q_{n_1}^2+R_T^2 \sin^2\beta_T^{(n_1)}}$, $
Q_{n_1} \approx D-R_T\cos\beta_T^{(n_1)} \times \cos\alpha_T^{(n_1)}$, $\xi_{n_2}=\sqrt{Q_{n_2}^2+R_R^2\sin^2\beta_R^{(n_2)}}$, $Q_{n_2} \approx D+R_R\cos\beta_R^{(n_2)}\cos\alpha_R^{(n_2)}$, $\xi_R^{(n_3)}=\frac{2a-Q_{n_3}}{\cos\beta_R^{(n_3)}}$, $\xi_T^{(n_3)}=\sqrt{Q_{n_3}^2+(\xi_R^{(n_3)})^2\sin^2\beta_R^{(n_3)}}$, and $Q_{n_3}=\frac{a^2+f^2+2af\cos\alpha_R^{(n_3)}}{a+f\cos\alpha_R^{(n_3)}}$.

Note that the azimuth/elevation angle of departure (AAoD/EAoD), (i.e., $\alpha_T^{(n_{i})}$, $\beta_T^{(n_{i})}$), and azimuth/elevation angle of arrival (AAoA/EAoA), (i.e.,  $\alpha_R^{(n_{i})}$, $\beta_R^{(n_{i})}$), are independent for double-bounced rays, while are correlated for single-bounced rays. According to geometric algorithms, for the single-bounced rays resulting from the two-sphere model, we can derive the relationship between the AoDs and AoAs as $\alpha_R^{(n_{1})} \approx \pi- \frac{R_T}{D} \sin\alpha_T^{(n_{1})}$, $\beta_R^{(n_{1})} \approx \arccos \big ( \frac{D-R_T \cos\beta_T^{(1)}\cos\alpha_T^{(1)}}{\xi_{n_1}} \big )$, and $\alpha_T^{(n_{2})} \approx \frac{R_R}{D} \sin\alpha_R^{(n_{2})}$, $\beta_T^{(n_{2})} \approx  \arccos \big ( \frac{D+R_R \cos \beta_R^{(2)} \cos \alpha_R^{(2)}}{\xi_{n_2}} \big )$. For the single-bounced rays resulting from elliptic-cylinder model, the angular relationship $\alpha_T^{(n_{3})}=\arcsin \big ( \frac{b^2 \sin \alpha_R^{(n_{3})}}{a^2 + f^2 + 2af \cos \alpha_R^{(n_{3})}} \big )$ and $\beta_T^{(n_3)}=\arccos \big [ \frac{a^2+f^2+2af \cos \alpha_R^{(n_3)}}{\left(a+f \cos \alpha_R^{(n_3)} \right) \xi_T^{(n_3)}} \big ]$ hold with $b=\sqrt{a^2-f^2}$ denoting the semi-minor axis of the elliptic-cylinder. The undefined $\beta^{(n_1)}_R$, $\beta^{(n_2)}_T$, and $\beta^{(n_3)}_T$ in Line 12 of the left column on Page 3 in \cite{XCheng10VTC} have been given here. 

\begin{figure} [t]%
\centering
\parbox{8.5cm}{%
\centering \hspace{-0.3cm} \includegraphics[scale=0.55]{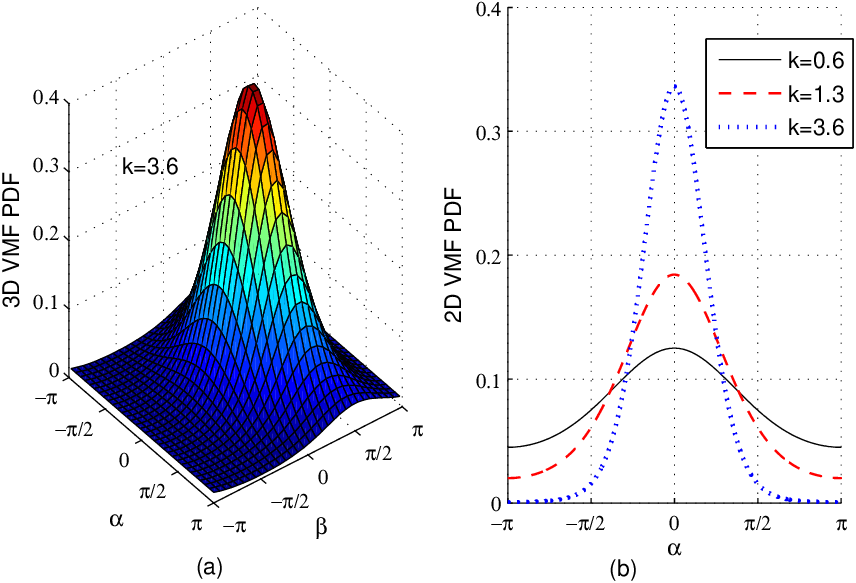} 
\caption{(a) The 3D VMF PDF ($\alpha_0=0^{\circ}$, $\beta_0=31.6^{\circ}$, $k=3.6$) and (b) 2D VMF PDF ($\alpha_0=0^{\circ}$, $\beta_0=31.6^{\circ}$, $\beta=0^{\circ}$, $k=0.6, 1.3, 3.6$).}%
}%
\end{figure}%

For the theoretical RS-GBSM, as the number of scatterers tends to infinity, the discrete AAoD $\alpha_T^{(n_{i})}$, EAoD $\beta_T^{(n_{i})}$, AAoA $\alpha_R^{(n_{i})}$, and EAoA $\beta_R^{(n_{i})}$ can be replaced by continuous random variables $\alpha_T^{({i})}$, $\beta_T^{({i})}$, $\alpha_R^{({i})}$, and $\beta_R^{({i})}$, respectively. In \cite{Mammasis08WCNC}, the assumption of 3D scattering has been validated. To jointly consider the impact of the azimuth and elevation angles on channel statistics, we use the VMF PDF to characterize the distribution of effective scatterers, which is defined as \cite{Mardia00} 
\begin{align}
f\left(\alpha, \beta\right) = \frac{k \cos \beta }{4\pi \sinh k} \times e^{ k \left [ \cos\beta_0\cos\beta \cos \left ( \alpha-\alpha_0 \right ) +\sin\beta_0\sin\beta \right ] }
\end{align}
where $\alpha, \beta\in{\left[-\pi,\pi\right)}$, $\alpha_0\in{\left[-\pi,\pi\right)}$ and $\beta_0\in{\left[-\pi,\pi\right)}$ account for the mean values of the azimuth angle $\alpha$ and elevation angle $\beta$, respectively, and $k$ ($k \geq 0$) is a real-valued parameter that controls the concentration of the distribution relative to the mean direction identified by $\alpha_0$ and $\beta_0$. 

To demonstrate the VMF distribution, we set the mean angles $\alpha_0=0^{\circ}$ and $\beta_0=31.6^{\circ}$ as an example, and plot the corresponding PDF in both 3D and 2D figures in Figs.~3 (a) and (b), respectively. Fig.~3 (a) shows the 3D VMF PDF with $k=3.6$. For the purpose of comparison, in Fig.~3 (b), we plot the 2D VMF PDF only for azimuth angle $\alpha$ with $\beta=0^{\circ}$ and different $k=0.6, 1.3, 3.6$. 
Fig.~3~(b) tells that the larger the value of $k$, the VMF PDF is more concentrated towards the mean direction. For $k \rightarrow 0$ the distribution is isotropic, while for $k \rightarrow \infty$ the distribution becomes extremely non-isotropic. For the high VTD scenario with many moving vehicles around the Tx and Rx, $k$ is small and the scatterer distribution approaches isotropic. Note that when the elevation angle $\beta=\beta_0=0^{\circ}$, the VMF PDF reduces to von Mises PDF, which has widely been applied as a scatterer distribution in 2D propagation environments \cite{Abdi02}. In this paper, for the angles of interest, i.e., the AAoD $\alpha_T^{(1)}$ and EAoD $\beta_T^{(1)}$ for the Tx sphere, the AAoA $\alpha_R^{(2)}$ and EAoA $\beta_R^{(2)}$ for the Rx sphere, and the AAoA $\alpha_R^{(3)}$ and EAoA $\beta_R^{(3)}$ for the elliptic-cylinder, the parameters ($\alpha_0$, $\beta_0$, and $k$) of the VMF PDF in (4) can be replaced by ($\alpha_{T0}^{(1)}$, $\beta_{T0}^{(1)}$, and $k^{(1)}$), ($\alpha_{R0}^{(2)}$, $\beta_{R0}^{(2)}$, and $k^{(2)}$), and ($\alpha_{R0}^{(3)}$, $\beta_{R0}^{(3)}$, and $k^{(3)}$), respectively. 

It is important to emphasize that the proposed model is adaptable to a wide variety of V2V propagation environments by adjusting important parameters, which are the Ricean factor $K$, energy-related parameters $\eta_{SB_i}$ and $\eta_{DB}$, and environment parameters $k^{(i)}$. In general, for a low VTD, the value of $K$ is large since the LoS component can bear a significant amount of power. In addition, the received scattered power is mainly from waves reflected by the stationary roadside environments described by the scatterers located on the elliptic-cylinder. The moving vehicles represented by the scatterers located on the two spheres are sparse and thus more likely to be single-bounced, rather than double-bounced. This indicates that $\eta_{SB_3} > \max \{\eta_{SB_1}, \eta_{SB_2}\} > \eta_{DB}$. For a high VTD, the value of $K$ is smaller than that in the low VTD scenario. Also, due to dense moving vehicles, the double-bounced rays of the two-sphere model bear more energy than single-bounced rays of the two-sphere and elliptic-cylinder models, i.e., $\eta_{DB} > \max \{\eta_{SB_1}, \eta_{SB_2}, \eta_{SB_3}\}$. Therefore, the consideration of the VTD can be well characterized by utilizing a combined two-sphere model and elliptic-cylinder model with the LoS component.

\subsection{Statistical properties of the 3D MIMO V2V RS-GBSM}
For the proposed 3D MIMO V2V theoretical RS-GBSM, statistical properties will be derived in this section, i.e., amplitude and phase PDFs, ST CF, Doppler PSD, envelope LCR, and AFD.

\subsubsection{Amplitude and phase PDFs}
Based on the proposed 3D theoretical RS-GBSM, the amplitude and phase processes can be expressed as $\zeta (t) = \left| h_{pq}(t)\right| $ and $\vartheta (t) = \arg \left\{ h_{pq}(t)\right\}$, respectively. According to the similar procedure in \cite{Patzold07}, the amplitude PDF of the 3D V2V reference model can be derived as  
\begin{align} \label{eq:amplitude_PDF}
p_\zeta (z) = \frac{z}{{\sigma _0^2}}{e^{ - \frac{{{z^2} + {K_0^2}}}{{2\sigma _0^2}}}} \textit{I}_0(\frac{{z{K_0}}}{{\sigma _0^2}})
\end{align}
where $z$ presents the amplitude variable, $K_0=\sqrt {\frac{K}{{K + 1}}}$ and $\textit{I}_0(\cdot)$ is the zeroth-order modified Bessel function of the first kind. 

In addition, the phase PDF of the reference model can be derived as
\begin{align} \nonumber \label{eq:phase_PDF}
p_\vartheta (\theta ) &= \frac{e^{ - \frac{K_0^2}{2\sigma _0^2}}}{2\pi}\left\{ 1 + \frac{K_0}{\sigma _0}\sqrt {\frac{\pi }{2}} \cos \left( {\theta - {\theta _K}} \right){e^{\frac{K_0^2{{\cos }^2}(\theta - {\theta _K})}{2\sigma _0^2}}} \right. \\ 
& \left.\times \left[ {1 + {\rm{erf}}\left( {\frac{{{K_0}\cos (\theta - {\theta _K})}}{{{\sigma _0}\sqrt 2 }}} \right)} \right] \right\}
\end{align}
where $\theta _K=\arg \left\{ h_{pq}^{LoS}(t)\right\}$. Due to the page limit, detailed derivations are omitted here.

\subsubsection{ST CF}
Under the wide-sense stationary (WSS) condition, the normalized ST CF between any two complex fading envelopes $h_{pq}\left(t\right)$ and $h_{p'q'}\left(t\right)$ is defined as \cite{XCheng10VTC} 
\begin{align} \label{eq:STCF}
\rho_{h_{pq}h_{p^\prime q^\prime}} \left( \tau \right) &= \frac{\mathbf{E} \left[h_{pq}(t)h_{p^\prime q^\prime}^{*}(t-\tau)\right]} {\sqrt{\mathbf{E} \left[ |h_{pq}^{^{}} (t) |^2 \right] \mathbf{E} \left[ \left|h_{p^\prime q^\prime}(t) \right|^2 \right]}} \\ \nonumber
&= \mathbf{E} \left[ h_{pq} \left( t \right) h_{p^\prime q^\prime}^{*} \left( t - \tau \right) \right] \left( K+1 \right)
\end{align}
where $(\cdot)^*$ denotes the complex conjugate operation and $\mathbf{E} [\cdot ]$ designates the statistical expectation operator. Substituting \eqref{eq:h_pq(t)} into \eqref{eq:STCF} and applying the corresponding VMF distribution, we can obtain the ST CF of the LoS, single-, and double-bounced components as follows: 

(a) In the case of the LoS component,
\begin{align} \label{eq:STCF_LoS}
\rho_{h_{pq}^{LoS}h_{p^\prime q^\prime}^{LoS}}\left(\tau\right)= K e^{\frac{j2\pi}{\lambda} A^{LoS}+j2\pi\tau( f_{T_{max}}\cos\gamma_T- f_{R_{max}}\cos\gamma_R) }
\end{align}
where $A^{LoS}=2D \cos\varphi_R\cos\theta_R$.

(b) In terms of the single-bounced components $SB_i$ ($i=1,2,3$) resulting from the Tx sphere, Rx sphere, and elliptic-cylinder, respectively,
\begin{align} \label{eq:STCF_SBi}
& \rho_{h_{pq}^{SB_{i}}h_{p^\prime q^\prime}^{SB_{i}}}\left(\tau\right) = \eta_{SB_{i}}\int_{-\pi}^{\pi} \int_{-\pi}^{\pi} \left [ e^{-\frac{j2\pi}{\lambda} A^{(i)}} \right.\\
& \left. \! \! \times e^{j2\pi\tau \left (f_{T_{max}}B^{(i)}+f_{R_{max}}C^{(i)} \right )} f(\alpha_{T/R}^{(i)}, \beta_{T/R}^{(i)}) \right ] d(\alpha_{T/R}^{(i)}, \beta_{T/R}^{(i)}) \nonumber
\end{align}
with $A^{(1)} = \delta_{T} \big[ \sin\beta_{T}^{(1)} \sin \varphi_{T} + \cos \beta_{T}^{(1)} \cos \varphi_{T} \cos \big( \theta_{T} - \alpha_{T}^{(1)}\big) \big] + \frac{\delta_{R}}{\xi_{n_{1}}} \big[ R_{T} \sin \beta_{T}^{(1)} \sin \varphi_{R} - Q_{n_{1}} \cos \varphi_{R} \cos \big( \theta_{R} - \alpha_{R}^{(1)} \big) \big]$, $B^{(i)} \! = \! \cos \big( \alpha_{T}^{(i)} \!-\! \gamma_T \big) \cos \big( \beta_T^{(i)} \big)$, $C^{(i)} \! = \! \cos \big( \alpha_{R}^{(i)} \!-\! \gamma_R \big) \times \cos \big( \beta_R^{(i)} \big)$, $A^{(2)} = \delta_{R} \big[ \! \sin \beta_{R}^{(2)} \sin \varphi_{R} + \cos \beta_{R}^{(2)} \cos \varphi_{R} \times \cos \left( \theta_{R} - \alpha_{R}^{(2)} \right) \! \big] + \frac{\delta_{T}}{\xi_{n_{2}}} \big[ R_{R} \sin \beta_{R}^{(2)} \sin \varphi_{T} + Q_{n_{2}} \cos \varphi_{T} \times \cos \left( \theta_{T} - \alpha_{T}^{(2)} \right) \! \big]$, \! $A^{(3)} = \frac{\delta_{T}}{\xi_T^{(n_3)}} \big[ \xi_R^{(n_3)} \sin \beta_{R}^{(3)} \sin \varphi_{T} + Q_{n_3} \times \cos \varphi_{T} \cos \left( \theta_{T} - \alpha_{T}^{(3)} \right) \! \big] \!+\! \delta_{R} \big[ \sin \beta_{R}^{(3)} \sin \varphi_{R} + \cos \beta_{R}^{(3)} \times \cos \varphi_{R} \cos \left( \theta_{R} - \alpha_{R}^{(3)} \right) \! \big]$, where the expressions of $\alpha_{R}^{(i)}$, $\beta_R^{(i)}$, $Q_{n_{i}}$, $\xi_{n_{1}}$, $\xi_{n_{2}}$, and $\xi^{n_{3}}_{T(R)}$ are given in Section II. A. Note that the subscripts $T$ and $R$ are applied to $i=1$ and $i=2,3$, respectively.

(c) In terms of the double-bounced component resulting from the Tx and Rx spheres,
\begin{align} \nonumber \label{eq:STCF_DB} 
&\rho_{h_{pq}^{DB}h_{p^\prime q^\prime}^{DB}}\left(\tau\right) = \rho_{p{p}'}^T(\tau) \rho_{q{q}'}^R(\tau) = \eta_{DB}\int_{-\pi}^{\pi} \int_{-\pi}^{\pi} \int_{-\pi}^{\pi} \int_{-\pi}^{\pi} \\ 
&\left[ e^{-\frac{j2\pi}{\lambda} A^{DB}} \cdot e^{j2\pi\tau \left (f_{T_{max}}B^{DB}+f_{R_{max}}C^{DB} \right )}\right. \\
& \times \left. f(\alpha_T^{(1)}, \beta_T^{(1)}) \cdot f(\alpha_R^{(2)}, \beta_R^{(2)}) \right] d(\alpha_{T}^{(1)}, \beta_{T}^{(1)}) d(\alpha_{R}^{(2)}, \beta_{R}^{(2)}) \nonumber
\end{align}
where $A^{DB} = \delta_{T} \big[ \sin \beta_{T}^{(1)} \sin \varphi_{T} + \cos \beta_{T}^{(1)} \cos \varphi_{T} \cos \big( \theta_{T} - \alpha_{T}^{(1)} \big) \big] + \delta_{R} \big[ \sin \beta_{R}^{(2)} \sin \varphi_{R} + \cos \beta_{R}^{(2)} \cos \varphi_{R} \cos \big( \theta_{R} - \alpha_{R}^{(2)} \big) \big]$, $B^{DB} = \cos \big( \alpha_{T}^{(1)} - \gamma_T \big) \cos \beta_T^{(1)}$, and $C^{DB} = \cos \big( \alpha_{R}^{(2)} - \gamma_R \big) \cos \beta_R^{(2)}$.

The normalized theoretical ST CF 
can be expressed as the summation of \eqref{eq:STCF_LoS} -- \eqref{eq:STCF_DB}, i.e.,
\begin{align} \nonumber \label{eq:STCF_all}
\rho_{h_{pq}h_{p^\prime q^\prime}}\left(\tau\right) &=\rho_{h_{pq}^{LoS}h_{p^\prime q^\prime}^{LoS}}\left(\tau\right) + \sum_{i=1}^{I} \rho_{h_{pq}^{SB_{i}}h_{p^\prime q^\prime}^{SB_{i}}}\left(\tau\right) \\ 
&+ \rho_{h_{pq}^{DB}h_{p^\prime q^\prime}^{DB}}\left(\tau\right).
\end{align}

\subsubsection{Doppler PSD}
Applying the Fourier transform to the ST CF, we can obtain the corresponding Doppler PSD as $S_{h_{pq}h_{p^{\prime} q^{\prime}}}(f_D)=\mathbf{F}\left\{\rho_{h_{pq}h_{p^\prime q^\prime}}(\tau)\right\}=\int_{-\infty}^{\infty} \rho_{h_{pq}h_{p^\prime q^\prime}}(\tau)e^{-j2\pi f_D\tau}d\tau$, where $f_D$ is the Doppler frequency. Substituting \eqref{eq:STCF_all} into the above equation, the Doppler PSD can be expressed as
\begin{align} \nonumber \label{eq:PSD}
S_{h_{pq}h_{p^\prime q^\prime}} \left( f_D \right) &= \mathbf{F} \! \left\{ \rho_{h_{pq}^{LoS} h_{p^\prime q^\prime}^{LoS}} \left( \tau \right) \right\} \! + \! \sum_{i=1}^{I} \mathbf{F} \! \left\{ \rho_{h_{pq}^{SB_i} h_{p^\prime q^\prime}^{SB_i}} \left( \tau \right) \right\} \\
&+ \mathbf{F} \left\{ \rho_{pp^\prime}^T \left( \tau \right) \right\} \odot \mathbf{F} \left\{ \rho_{qq^\prime}^R \left( \tau \right) \right\}
\end{align}
where $\odot$ denotes the convolution and $\mathbf{F} \{ \cdot \}$ indicates the Fourier transform.

\subsubsection{Envelope LCR and AFD}
The LCR at a specified level $r$, $L(r)$, is defined as the rate at which the signal envelope crosses level $r$ in the positive/negative going direction. Using the traditional PDF-based method \cite{Zajic08ICC}, we derive the expression of the LCR for V2V channels as 
\begin{align} \nonumber \label{eq:LCR}
 L(r)&=\frac{2r\sqrt{K+1}}{\pi^{3/2}}\sqrt{\frac{b_2}{b_0}-\frac{b^2_1}{b^2_0}} \times e^{-K-(K+1)r^2} \\ 
&\times \int_{0}^{\pi/2} \cosh \left (2\sqrt{K(K+1)} \cdot r\cos \theta \right ) \\ 
&\times \left [ e^{-(\chi\sin \theta)^2 }+\sqrt{\pi}\chi\sin \theta \cdot {\mbox{erf}}(\chi\sin \theta) \right ] d\theta \nonumber
\end{align}
where $\cosh(\cdot)$ is the hyperbolic cosine function, $\mbox {erf}(\cdot)$ is the error function, and $\chi = \sqrt{\frac {K b_1^2} {\left( b_0 b_2 - b_1^2 \right)}}$. Finally, parameters $b_0$, $b_1$, and $b_2$ are defined as
\begin{align} \label{eq:b0}
b_0 \buildrel\triangle\over = \textit{E} \left[ h_{pq}^{In}(t)^2 \right]=\textit{E} \left[ h_{pq}^{Qu}(t)^2 \right]
\end{align}
\begin{align} \label{eq:b1}
b_1 \buildrel\triangle\over = \textit{E} \left[ h_{pq}^{In}(t)\dot{h}_{pq}^{Qu}(t) \right]=\textit{E} \left[ h_{pq}^{Qu}(t)\dot{h}_{pq}^{In}(t) \right]
\end{align}
\begin{align} \label{eq:b2}
b_2 \buildrel\triangle\over = \textit{E} \left[ \dot{h}_{pq}^{In}(t)^2 \right]=\textit{E} \left[ \dot{h}_{pq}^{Qu}(t)^2 \right]
\end{align}
where $h_{pq}^{In}(t)$ and $h_{pq}^{Qu}(t)$ denote the in-phase and quadrature components of the complex fading envelope $h_{pq}(t)$, and $\dot{h}_{pq}^{In}(t)$ and $\dot{h}_{pq}^{Qu}(t)$ denote the first derivative of $h_{pq}^{In}(t)$ and $h_{pq}^{Qu}(t)$, respectively. 
By substituting \eqref{eq:h_pq(t)} into \eqref{eq:b0}, the parameter $b_0$ becomes
\begin{align}
b_0 = \sum_{i=1}^{I} b_0^{SB_i}+b_0^{DB}=\frac{1}{2(K+1)}
\end{align} 
where
\begin{subequations}
\begin{align} \nonumber \label{eq:b0SBi}
b_0^{SB_i} &= \frac{\eta_{SB_i}}{2(K+1)}\int_{-\pi}^{\pi}\int_{-\pi}^{\pi} f(\alpha_T^{(i)}, \beta_T^{(i)}) d(\alpha_T^{(i)}, \beta_T^{(i)}) \\
&=\frac{\eta_{SB_i}}{2(K+1)}
\end{align}
\begin{align} \label{eq:b0DB}
b_0^{DB} &= \frac{\eta_{DB}}{2(K+1)}\int_{-\pi}^{\pi}\int_{-\pi}^{\pi} f(\alpha_T^{(1)}, \beta_T^{(1)}) d(\alpha_T^{(1)}, \beta_T^{(1)}) \nonumber \\
&\! \! \! \! \! \! \! \! \! \! \! \times \int_{-\pi}^{\pi}\int_{-\pi}^{\pi} f(\alpha_R^{(2)}, \beta_R^{(2)}) d(\alpha_R^{(2)}, \beta_R^{(2)}) =\frac{\eta_{DB}}{2(K+1)}.
\end{align}
\end{subequations}
Similarly, by substituting \eqref{eq:h_pq(t)} into \eqref{eq:b1} and \eqref{eq:b2}, the parameters $b_1$ and $b_2$ become
\begin{align}
b_m = \sum_{i=1}^{I} b_m^{SB_i} + b_m^{DB},
\end{align} 
where $m \in \{1, 2\}$ and 
\begin{subequations}
\begin{align} \nonumber \label{eq:bmSBi}
b_m^{SB_i} &= \frac{\eta_{SB_i}}{2(K+1)}(2\pi)^m \int_{-\pi}^{\pi}\int_{-\pi}^{\pi} f(\alpha_R^{(i)}, \beta_R^{(i)}) \\ 
& \times \left [ f_{T_{max}}\cos\left ( \alpha_T^{(i)}-\gamma_T \right )\cos \beta_T^{(i)} \right ]^m \\ \nonumber 
& \times \left [ f_{R_{max}}\cos\left ( \alpha_R^{(i)}-\gamma_R \right )\cos \beta_R^{(i)} \right ]^m d(\alpha_R^{(i)}, \beta_R^{(i)})
\end{align}
\begin{align} \nonumber \label{eq:bmDB}
b_m^{DB} &= \frac{\eta_{DB}}{2(K+1)}(2\pi)^m \int_{-\pi}^{\pi}\int_{-\pi}^{\pi} f(\alpha_T^{(1)}, \beta_T^{(1)}) \\ \nonumber
& \times \left [ f_{T_{max}}\cos\left ( \alpha_T^{(1)}-\gamma_T \right )\cos \beta_T^{(1)} \right ]^m d(\alpha_T^{(1)}, \beta_T^{(1)}) \\ 
& \times \int_{-\pi}^{\pi}\int_{-\pi}^{\pi} f(\alpha_R^{(2)}, \beta_R^{(2)}) \\ 
& \times \left [ f_{R_{max}}\cos\left ( \alpha_R^{(2)}-\gamma_R \right )\cos \beta_R^{(2)} \right ]^m d(\alpha_R^{(2)}, \beta_R^{(2)}). \nonumber
\end{align}
\end{subequations}

The AFD, $T(r)$, is defined as the average time over which the signal envelope, $\left | h_{pq}(t) \right |$, remains below a certain level $r$. In the proposed 3D RS-GBSM, the AFD can be written as \cite{Stuber01}
\begin{align} \label{eq:AFD}
T(r)=\frac{1-\mathbf{Q} \left( \sqrt{2K},\sqrt{2(K+1)r^2} \right)}{L(r)}
\end{align}
where $\mathbf{Q} \left(\ \cdot \ ,\ \cdot \ \right)$ is the Marcum Q function.

\section{The 3D SoS Simulation Model for MIMO V2V Channels} 

Based on the proposed 3D theoretical RS-GBSM described in Section~II, the corresponding SoS simulation model can be further developed by using finite numbers of scatterers or sinusoids $N_1$, $N_2$, and $N_3$. According to \eqref{eq:h_pq(t)} -- (2c), the SoS simulation model for the link $T_p \to T_q$ can be expressed as 
\begin{align}  \label{eq:h_pq(t)_sim}
\hat{h}_{pq}\left(t\right) &= \hat{h}_{pq}^{LoS}\left(t\right) + \sum_{i=1}^{I}\hat{h}_{pq}^{SB_i}\left(t\right) + \hat{h}_{pq}^{DB}\left(t\right) 
\end{align}
where
\begin{subequations}
\begin{align} \nonumber \label{eq:h_pq(t)_LoS_sim}
\hat{h}_{pq}^{LoS}(t) &= \sqrt{\frac{K}{K+1}}e^{-j2\pi f_c \tau_{pq}} \\
& \times e^{j2\pi f_{T_{max}} t \cos\left(\alpha_T^{LoS}-\gamma_T\right)\cos{\beta_T^{LoS}}} \\ \nonumber
& \times e^{j2\pi f_{R_{max}} t \cos\left(\alpha_R^{LoS}-\gamma_R\right)\cos{\beta_R^{LoS}}}
\end{align}
\begin{align} \nonumber \label{eq:h_pq(t)_SB_sim}
\hat{h}_{pq}^{SB_i}\left(t\right) &= \sqrt{\frac{\eta_{SB_{i}}}{K+1}}\sum_{n_{i}=1}^{N_{i}}\frac{1}{\sqrt{N_{i}}}e^{j\left(\psi_{n_{i}}-2\pi f_c \tau_{pq,n_{i}}\right)} \\
& \times e^{j2\pi f_{T_{max}} t \cos\left(\alpha_T^{(n_{i})}-\gamma_T \right)\cos{\beta_T^{(n_i)}}} \\ \nonumber
& \times e^{j2\pi f_{R_{max}} t \cos \left(\alpha_R^{(n_{i})}-\gamma_R \right)\cos{\beta_R^{(n_i)}}}
\end{align}
\begin{align} \nonumber  \label{eq:h_pq(t)_DB_sim}
\hat{h}_{pq}^{DB}\left(t\right) &=\sqrt{\frac{\eta_{DB}}{K+1}}\sum_{n_{1},n_{2}=1}^{N_{1},N_{2}}\frac{1}{\sqrt{N_{1}N_{2}}}e^{j\left(\psi_{n_{1},n_{2}}-2\pi f_c \tau_{pq,n_{1},n_{2}}\right)} \\ 
& \times e^{j2\pi f_{T_{max}} t \cos\left(\alpha_T^{(n_{1})}-\gamma_T\right)\cos{\beta_T^{(n_1)}}} \\ \nonumber
& \times e^{j2\pi f_{R_{max}} t \cos\left(\alpha_R^{(n_{2})}-\gamma_R\right)\cos{\beta_R^{(n_2)}}}.
\end{align}
\end{subequations}

It is clear that the unknown simulation model parameters to be determined are only the discrete AoDs and AoAs, while the remaining parameters are identical to those of the theoretical model. Our task is thus to determine the discrete AAoDs ($\alpha_T^{(n_1)}$, $\alpha_T^{(n_2)}$, $\alpha_T^{(n_3)}$), EAoDs ($\beta_T^{(n_1)}$, $\beta_T^{(n_2)}$, $\beta_T^{(n_3)}$), AAoAs ($\alpha_R^{(n_1)}$, $\alpha_R^{(n_2)}$, $\alpha_R^{(n_3)}$), and EAoAs ($\beta_R^{(n_1)}$, $\beta_R^{(n_2)}$, $\beta_R^{(n_3)}$) for the simulation model. Furthermore, there are actually correlations between AoDs and AoAs for the single-bounce case. Therefore, we only need to determine the discrete sets of $\left\{\alpha_T^{(n_1)}, \beta_T^{(n_1)}\right\}_{n_1=1}^{N_1}$, $\left\{\alpha_R^{(n_2)}, \beta_R^{(n_2)}\right\}_{n_2=1}^{N_2}$, and $\left\{\alpha_R^{(n_3)}, \beta_R^{(n_3)}\right\}_{n_3=1}^{N_3}$. In \cite{Patzold12}, different parameter computation methods have been introduced. 
In general, there are three widely adopted methods, i.e., Extended Method of Exact Doppler Spread (EMEDS), Modified Method of Equal Areas (MMEA), and Lp-Norm method (LPNM). The EMEDS is especially recommended for isotropic scattering. However, all the above methods are only valid for 2D horizontal models. To jointly calculate the azimuth and elevation angles, we propose a novel parameter computation method that can be applied to our 3D channel models. The method is named as MEV, which is developed from MMEA \cite{Xcheng13CL}.

\subsection{MEV for parameterization of the proposed SoS simulation model}
As we mentioned before, the VMF distribution is adopted in order to jointly consider the impact of the azimuth and elevation angles on channel statistics. Furthermore, the cumulative distribution function (CDF) of $\alpha$ and $\beta$, i.e., the double integral of the 3D VMF PDF, denotes the volume of Fig.~3 (a). The idea of MEV is designed to select the set of $\left\{\alpha^{(n_i)}, \beta^{(n_i)} \right\} _{n_i=1}^{N_i}$ in such a manner that the volume of the VMF PDF $f(\alpha, \beta)$ in different ranges of $\left \{ \alpha ^{(n_i-1)},\beta ^{(n_i-1)} \right \} \leqslant \left \{ \alpha ,\beta  \right \} < \left \{ \alpha ^{(n_i)},\beta ^{(n_i)} \right \}$ are equal
to each other with the initial condition $\int_{-\pi}^{\alpha^{(1)}} \hspace{-0.2cm} \int_{-\pi}^{\beta ^{(1)}} \hspace{-0.2cm} f(\alpha ,\beta )d\alpha d\beta =\frac{1-1/4}{N_i}$. The application of the MEV to the 3D V2V channel model requires the joint computation of the discrete model parameters, i.e., $\left\{\alpha_T^{(n_1)}, \beta_T^{(n_1)}\right\}_{n_1=1}^{N_1}$, $\left\{\alpha_R^{(n_2)}, \beta_R^{(n_2)}\right\}_{n_2=1}^{N_2}$, and $\left\{\alpha_R^{(n_3)}, \beta_R^{(n_3)}\right\}_{n_3=1}^{N_3}$. 
In the following, we will derive the MEV that has the ability to meet the two accuracy-efficiency design criteria \cite{Xcheng13CL} for 3D scattering MIMO V2V channels with the joint VMF distribution. Using the design of the AAoDs $\left\{\alpha_T^{(n_1)}\right\}_{n_1=1}^{N_1}$ and EAoDs $\left\{\beta_T^{(n_1)}\right\}_{n_1=1}^{N_1}$ as an example, the MEV includes the following three steps:

Step 1: Define a pair of random variables, i.e., $\acute{\alpha_T}^{(n_1)}\in\left[\alpha_{T0}^{(1)}-\pi,\alpha_{T0}^{(1)}+\pi\right)$ and $\acute{\beta_T}^{(n_1)}\in\left[\beta_{T0}^{(1)}-\pi,\beta_{T0}^{(1)}+\pi \right)$. They follow the VMF distribution having the same $\alpha_{T0}^{(1)}$, $\beta_{T0}^{(1)}$, and $k_1$.

Step 2: Temporarily design the proper set of $\left\{\acute{\alpha_T}^{(n_1)}\right\}_{n_1=1}^{N_1}$ and $\left\{\acute{\beta_T}^{(n_1)}\right\}_{n_1=1}^{N_1}$, as $\acute{\alpha_T}^{(n_1)},\acute{\beta_T}^{(n_1)} :=F^{-1}_{\alpha / \beta } \left(\frac{n_1-1/4}{N_1}\right)$, where $F^{-1}_{\alpha / \beta }(\cdot)$ denotes the inverse function of the VMF CDF derived from VMF PDF for $\acute{\alpha_T}^{(n_1)}$ and $\acute{\beta_T}^{(n_1)}$.

Step 3: Obtain the desired set of $\left\{\alpha_T^{(n_1)}\right\}_{n_1=1}^{N_1}$ and $\left\{\beta_T^{(n_1)}\right\}_{n_1=1}^{N_1}$ by mapping $\left\{\acute{\alpha_T}^{(n_1)}\right\}_{n_1=1}^{N_1}$ and $\left\{\acute{\beta_T}^{(n_1)}\right\}_{n_1=1}^{N_1}$ into the range of $\left[-\pi, \pi \right)$, respectively.


Consequently, the jointly calculated AAoDs and EAoDs, $\left\{\alpha_T^{(n_1)}, \beta_T^{(n_1)}\right\}_{n_1=1}^{N_1}$ are obtained. Similarly, AAoAs $\left\{\alpha_R^{(n_2)}\right\}_{n_2=1}^{N_2}$ and $\left\{\alpha_R^{(n_3)}\right\}_{n_3=1}^{N_3}$ and EAoAs $\left\{\beta_R^{(n_2)}\right\}_{n_2=1}^{N_2}$ and $\left\{\beta_R^{(n_3)}\right\}_{n_3=1}^{N_3}$ can be obtained by following the same procedure. 

\subsection{Statistical properties of the proposed SoS simulation model}
Based on our 3D MIMO V2V theoretical RS-GBSM and its statistical properties, it is achievable to derive the corresponding statistical properties for the SoS simulation model. As the detailed derivations have been explained in Section II.B, those of the corresponding simulation model with similar derivations are only briefly explained. Applying the discrete model parameters to \eqref{eq:amplitude_PDF}, \eqref{eq:phase_PDF}, \eqref{eq:STCF_all}, \eqref{eq:PSD}, \eqref{eq:LCR}, and \eqref{eq:AFD}, we have the corresponding statistical properties for the SoS simulation model as follows:

\subsubsection{Amplitude and Phase PDFs}
The amplitude and phase processes of the SoS simulation model can be expressed as $\hat{\zeta} (t) = \left| \hat{h}_{pq}(t)\right|$ and $\hat{\vartheta} (t) = \arg \left\{ \hat{h}_{pq}(t)\right\}$, respectively. Still using the similar procedure in \cite{Patzold07}, the amplitude PDF of the SoS simulation model can be derived as  
\begin{align} \label{eq:amplitude_PDF_sim}
p_{\hat{\zeta}} (z) &= 4{\pi ^2 }z\int_0^\infty \left[ \prod\limits_{{n_1} = 1}^{{N_1}} {{J_0}\left( {2\pi \left| {G_{SB_1}} \right|x} \right)} \right. \\ \nonumber
& \times \prod\limits_{{n_2} = 1}^{{N_2}} {{J_0}\left( {2\pi \left| {G_{SB_2}} \right|x} \right)} \times \prod\limits_{{n_3} = 1}^{{N_3}} {{J_0}\left( {2\pi \left| {G_{SB_3}} \right|x} \right)} \\ 
& \left. { \times \prod\limits_{{n_1},{n_2} = 1}^{{N_1},{N_2}} {{J_0}\left( {2\pi \left| {G_{DB}} \right|x} \right)} } \right] {J_0}\left( {2\pi zx} \right){J_0}\left( {2\pi {K_0}x} \right)xdx \nonumber
\end{align}
where $G_{SB_i}=\sqrt {\frac{\eta _{SB_i}}{{N_i}(K + 1)}}$ $(i=1,2,3)$, and $G_{DB}=\sqrt {\frac{\eta _{DB}}{{N_1 N_2}(K + 1)}}$.

In addition, the phase PDF of the SoS simulation model can be derived as
\begin{align} \nonumber \label{eq:phase_PDF_sim}
p_{\hat{\vartheta}} (\theta ) &= 2\pi \int_0^\infty  \int_0^\infty  \left[ \prod\limits_{{n_1} = 1}^{{N_1}} {{J_0}\left( {2\pi \left| {G_{SB_1}} \right|x} \right)} \right. \\ \nonumber
& \times \prod\limits_{{n_2} = 1}^{{N_2}} {{J_0}\left( {2\pi \left| {G_{SB_2}} \right|x} \right)} \times \prod\limits_{{n_3} = 1}^{{N_3}} {{J_0}\left( {2\pi \left| {G_{SB_3}} \right|x} \right)} \\
& \left. { \times \prod\limits_{{n_1},{n_2} = 1}^{{N_1},{N_2}} {{J_0}\left( {2\pi \left| {G_{DB}} \right|x} \right)} } \right] \\ \nonumber
& \times {J_0}\left( {2\pi x \sqrt {{z^2} + K_0^2 - 2z{K_0}\cos \left( {\theta  - {\theta _K}} \right)} } \right)xzdxdz.
\end{align}

\subsubsection{ST CF}
As we should represent the spatial components, here we rewrite the ST CF as 
\begin{align} \label{eq:STCF_sim}
&\hat{\rho}_{h_{pq}h_{p^\prime q^\prime}} \left ( \delta _T, \delta _R, \tau \right ) = \hat{\rho}_{h_{pq}^{LoS}h_{p^\prime q^\prime}^{LoS}} \left ( \delta _T, \delta _R, \tau \right ) \\ \nonumber
&+ \sum_{i=1}^{I} \hat{\rho}_{h_{pq}^{SB_{i}}h_{p^\prime q^\prime}^{SB_{i}}} \left ( \delta _T, \delta _R, \tau \right ) + \hat{\rho}_{h_{pq}^{DB}h_{p^\prime q^\prime}^{DB}} \left ( \delta _T, \delta _R, \tau \right ).
\end{align}

(a) In the case of the LoS component,
\begin{align} \label{eq:STCF_sim_LoS}
&\hat{\rho}_{h_{pq}^{LoS}h_{p^\prime q^\prime}^{LoS}}\left ( \delta _T, \delta _R, \tau \right ) = K e^{\frac{j2\pi}{\lambda} A^{LoS}} \\ \nonumber
&\times e^{j2\pi\tau( f_{T_{max}}\cos\gamma_T- f_{R_{max}}\cos\gamma_R) }.
\end{align}
Please note that the LoS ST CF of the SoS simulation model is identical to that of the 3D theoretical RS-GBSM. 

(b) In terms of the single-bounced components $SB_i$ $(i=1,2,3)$ resulting from the Tx sphere, Rx sphere, and elliptic-cylinder, respectively,
\begin{align} \label{eq:STCF_sim_SBi}
&\hat{\rho}_{h_{pq}^{SB_{i}}h_{p^\prime q^\prime}^{SB_{i}}}\left ( \delta _T, \delta _R, \tau \right ) = \frac{\eta_{SB_{i}}}{N_i} \sum_{n_i=1}^{N_i} e^{\frac{j2\pi}{\lambda} A^{(i)}} \\ \nonumber
&\times e^{j2\pi\tau \left (f_{T_{max}}B^{(i)}+f_{R_{max}}C^{(i)} \right )}.
\end{align}




(c) In terms of the double-bounced component resulting from the Tx and Rx spheres,
\begin{align} \nonumber \label{eq:STCF_sim_DB}
&\hat{\rho}_{h_{pq}^{DB}h_{p^\prime q^\prime}^{DB}}\left ( \delta _T, \delta _R, \tau \right ) = \hat{\rho}_{p{p}'}^T( \delta _T, \tau) \hat{\rho}_{q{q}'}^R( \delta _R, \tau) \\ \nonumber
& = \eta_{DB} \times \frac{1}{N_1} \sum_{n_1}^{N_1} e^{\frac{j2\pi}{\lambda} A^{DB_T}} \times e^{j2\pi\tau  f_{T_{max}}B^{DB} } \\  
& \times \frac{1}{N_2} \sum_{n_2}^{N_2} e^{\frac{j2\pi}{\lambda} A^{DB_R}} \times e^{j2\pi\tau f_{R_{max}}C^{DB} } 
\end{align}
where $A^{DB_T} = \delta_{T} \big[ \sin \beta_{T}^{(1)} \sin \varphi_{T} + \cos \beta_{T}^{(1)} \cos \varphi_{T} \cos \big(\theta_{T} - \alpha_{T}^{(1)} \big) \big]$, $A^{DB_R} = \delta_{R} \big[ \sin \beta_{R}^{(2)} \sin \varphi_{R} + \cos \beta_{R}^{(2)} \cos \varphi_{R} \times \cos \big(\theta_{R} - \alpha_{R}^{(2)} \big) \big]$, and $A^{LoS}$, $A^{(i)}$, $B^{(i)}$, $C^{(i)}$, $B^{DB}$, and $C^{DB}$ have been given in Section II.~B. 

\subsubsection{Doppler PSD}
The Doppler PSD of the SoS simulation model can be expressed as 
\begin{align} \nonumber \label{eq:PSD_sim}
\hat{S}_{h_{pq}h_{p^\prime q^\prime}}\left (f_D\right ) \! &= \mathbf{F} \! \left\{\hat{\rho}_{h_{pq}^{LoS}h_{p^\prime q^\prime}^{LoS}} \left(\tau\right)\right\} \!+ \! \sum_{i=1}^{I}\mathbf{F} \!\left\{\hat{\rho}_{h_{pq}^{SB_i}h_{p^\prime q^\prime}^{SB_i}} \left(\tau\right)\right\} \\
&+\mathbf{F}\left\{\hat{\rho}_{pp^\prime}^T\left(\tau\right)\right\}\odot\mathbf{F} \left\{\hat{\rho}_{qq^\prime}^R\left(\tau\right)\right\}.
\end{align}

\vspace{-0cm}
\subsubsection{Envelope LCR and AFD}
Similarly, according to \eqref{eq:LCR}, the envelope LCR of the SoS simulation model, $\hat{L}(r)$, can be derived as
\begin{align} \nonumber  \label{eq:LCR_sim}
 \hat{L}(r)&=\frac{2r\sqrt{K+1}}{\pi^{3/2}}\sqrt{\frac{\hat{b}_2}{\hat{b}_0}-\frac{\hat{b}^2_1}{\hat{b}^2_0}} \\ \nonumber 
&\times e^{-K-(K+1)r^2} \times \int_{0}^{\pi/2} \cosh \left (2\sqrt{K(K+1)} \cdot r\cos \theta \right ) \\ 
&\times \left [ e^{-(\hat{\chi}\sin \theta)^2 }+\sqrt{\pi}\hat{\chi}\sin \theta \cdot {\mbox{erf}}(\hat{\chi}\sin \theta) \right ] d\theta
\end{align}
with $\hat{\chi}=\sqrt{ \frac{K \hat{b}_1^2}{\hat{b}_0 \hat{b}_2 - \hat{b}_1^2}}$, $\hat{b}_0=\frac{1}{2(K+1)}$, and $\hat{b}_m = \sum_{i=1}^{I} \hat{b}_m^{SB_i}+\hat{b}_m^{DB}$ $(m=1,2)$, where
\begin{subequations}
\begin{align} \nonumber \label{eq:bmSBi_sim}
\hat{b}_m^{SB_i} &= \frac{\eta_{SB_i}}{2(K+1)}(2\pi)^m \\ 
& \times \frac{1}{N_i} \sum_{n_i=1}^{N_i} \left [ f_{Tmax}\cos\left ( \alpha_T^{(n_i)}-\gamma_T \right )\cos \beta_T^{(n_i)} \right. \\
& \times \left. f_{Rmax}\cos\left ( \alpha_R^{(n_i)}-\gamma_R \right )\cos \beta_R^{(n_i)} \right ]^m \nonumber 
\end{align}
\begin{align} \label{eq:bmDB_sim}
\hat{b}_m^{DB} &= \frac{\eta_{DB}}{K+1}(2\pi)^m \\
& \times \frac{1}{N_1} \sum_{n_1=1}^{N_1} \left [ f_{Tmax}\cos\left ( \alpha_T^{(n_1)}-\gamma_T \right )\cos \beta_T^{(n_1)} \right ]^m \nonumber \\
& \times \frac{1}{N_2} \sum_{n_2=1}^{N_2}\left [ f_{Rmax}\cos\left ( \alpha_R^{(n_2)}-\gamma_R \right )\cos \beta_R^{(n_2)} \right ]^m. \nonumber
\end{align}
\end{subequations}

Similarly, according to \eqref{eq:AFD}, the envelope AFD of the SoS simulation model, $\hat{T}(r)$, can be expressed as 
\begin{align} \label{eq:AFD_sim}
\hat{T}(r)=\frac{1-\mathbf{Q} \left( \sqrt{2K},\sqrt{2(K+1)r^2} \right)}{\hat{L}(r)}.
\end{align}


\begin{table*} [t]
\caption{Key parameters of different VTD scenarios.}
\centering
\begin{tabular}{|c|c|c|c|c|c|c|c|c|}
\hline
& \small $K$ & \small $\eta_{SB_1}$ & \small $\eta_{SB_2}$ & \small $\eta_{SB_3}$ & \small $\eta_{DB}$ & \small $k^{(1)}$ & \small $k^{(2)}$ & \small $k^{(3)}$ \\
\hline
\small Low VTD & \small 3.786 & \small 0.335 & \small 0.203 & \small 0.411 & \small 0.051 & \small 9.6 & \small 3.6 & \small 11.5 \\
\hline
\small High VTD & \small 0.156 & \small 0.126 & \small 0.126 & \small 0.063 & \small 0.685 & \small 0.6 & \small 1.3 & \small 11.5 \\
\hline
\end{tabular}
\end{table*}

\section{Simulation Results and Analysis}
In this section, we investigate both the 3D and 2D models in detail for each statistical property. Based on measured scenarios in \cite{Acosta07}, the following main parameters were chosen for our simulations: $f_c=5.9$~GHz, $D=300$~m, $f_{Tmax}=f_{Rmax}=570$~Hz, $a=180$~m, $R_T=R_R=15$~m, $\gamma_T=\gamma_R=0^{\circ}$, $\varphi_T=\varphi_R=45^{\circ}$, $\theta_T=\theta_R=45^{\circ}$, $\alpha_{T0}^{(1)}=21.7^{\circ}$, $\beta_{T0}^{(1)}=6.7^{\circ}$, $\alpha_{R0}^{(2)}=147.8^{\circ}$, $\beta_{R0}^{(2)}=17.2^{\circ}$, $\alpha_{R0}^{(3)}=171.6^{\circ}$, and $\beta_{R0}^{(3)}=31.6^{\circ}$. Considering the constraints of the Ricean factor and power-related parameters in \cite{XCheng10VTC}, we have $k^{(1)}=9.6$, $k^{(2)}=3.6$, $k^{(3)}=11.5$, $K=3.786$, $\eta_{SB_1}=0.335$, $\eta_{SB_2}=0.203$, $\eta_{SB_3}=0.411$, and $\eta_{DB}=0.051$ for low VTD scenario. For high VTD scenario, we have $k^{(1)}=0.6$, $k^{(2)}=1.3$, $k^{(3)}=11.5$, $K=0.156$, $\eta_{SB_1}=0.126$, $\eta_{SB_2}=0.126$, $\eta_{SB_3}=0.063$, and $\eta_{DB}=0.685$. Please note that $k^{(3)}=11.5$ for both low VTD and high VTD scenarios are applied. Table II summarizes key parameters adopted by low and high VTD scenarios. The environment-related parameters $k^{(1)}$, $k^{(2)}$, and $k^{(3)}$ are related to the distribution of scatterers (normally, the smaller values of $k^{(1)}$ and $k^{(2)}$ the more dense moving vehicles/scatterers, i.e., the higher VTD). In both high and low VTDs, $k^{(3)}$ is large as the scatterers reflected from static roadsides are normally concentrated. Also, Ricean factor $K$ is small in higher VTD, as the LoS component does not have dominant power. The reason is that dense vehicles (i.e., more vehicles/obstacles between Tx and Rx) on the road result in less likelihood of strong LoS components. For the SoS simulation model, we must first choose adequate values for the numbers of discrete scatterers $N_1$, $N_2$, and $N_3$. Based on our own simulation experiences and suggested by \cite{Patzold12}, a reasonable values for $N_i$ can be 40, which can be considered as a good trade-off between realization complexity and accuracy. Certainly, if we simulate rigorous channels, e.g., very high VTD, the number of effective scatterers can be increased to improve the performance of the channel simulator. In addition, when $\beta_{T}^{(n1)} = \beta_{R}^{(n2)} = \beta_{R}^{(n3)} = 0^{\circ}$, the proposed 3D model will be reduced to a 2D two-ring and elliptic model. The impact of elevation angle is evaluated in this section by comparing between the 3D and 2D models in terms of their statistical properties. 

\subsection{Amplitude and phase PDFs}

Figs.~4 and 5 show the amplitude and phase PDFs, respectively, for the 3D reference model, 3D simulation model with $N_1=N_2=N_3=40$, and 3D simulation results for both low and high VTD scenarios. Note that the simulation results were obtained from the channel coefficients generated by the proposed channel simulator. It is clear that both amplitude and phase PDFs of the simulation model, i.e., \eqref{eq:amplitude_PDF_sim} and \eqref{eq:phase_PDF_sim}, respectively, are completely determined by the number of scatterers $N_i$, the gains $G_{SB_i}$ and $G_{DB}$, and LoS amplitude $K_0$, whereas other model parameters have no influence at all. 
In addition, Figs.~4 and 5 demonstrate that the choice of $N_1=N_2=N_3=40$ is sufficient to obtain an excellent agreement between the simulation model and reference model in both low and high VTD scenarios.

\begin{figure} [t]%
\centering
\parbox{8.5cm}{%
\centering \includegraphics[scale=0.62]{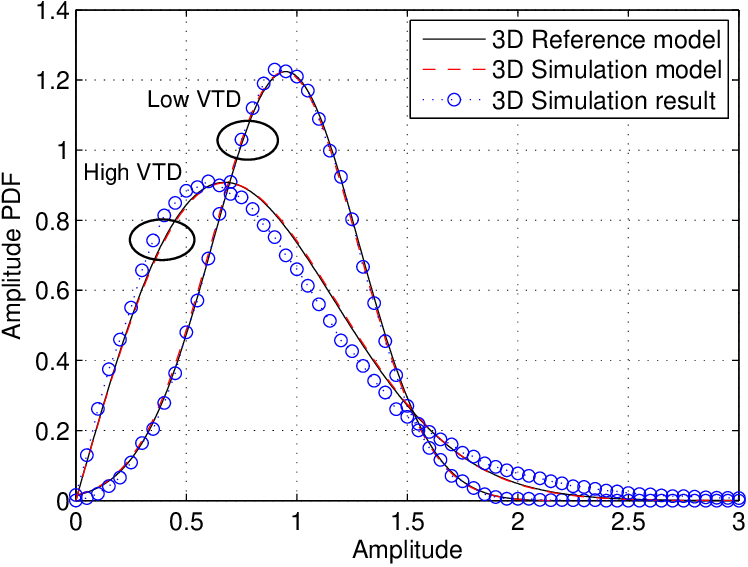} 
\vspace*{-0.2cm} 
\caption{The amplitude PDFs for the 3D reference model, 3D simulation model, and 3D simulation result ($\delta_T = \delta_R=0$, $\beta_{T0}^{(1)}= 6.7^{\circ}$, $\beta_{R0}^{(2)} = 17.2^{\circ}$, $\beta_{R0}^{(3)} = 31.6^{\circ}$).}%
}%
\quad

\parbox{8.5cm}{%
\centering \vspace*{0.5cm} \includegraphics[scale=0.62]{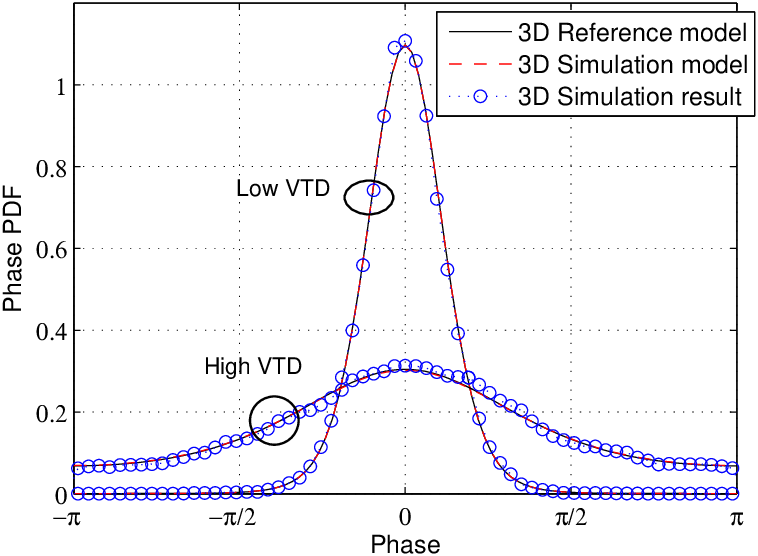} 
\vspace*{-0.2cm}
\caption{The phase PDFs for the 3D reference model, 3D simulation model, and 3D simulation result ($\delta_T = \delta_R=0$, $\beta_{T0}^{(1)}= 6.7^{\circ}$, $\beta_{R0}^{(2)} = 17.2^{\circ}$, $\beta_{R0}^{(3)} = 31.6^{\circ}$).}%
}%
\end{figure}%

\subsection{Temporal autocorrelation function}
We investigate the temporal autocorrelation function (ACF), which can be derived from the ST CF \eqref{eq:STCF_sim} by setting $d_T=d_R=0$. Therefore, the temporal ACF can be expressed as 
\begin{align}
\hat{\rho}_{h_{pq}h_{p^\prime q^\prime}}\left(\tau\right)=\hat{\rho}_{h_{pq}h_{p^\prime q^\prime}}\left(0,0,\tau\right).
\end{align}

Fig.~6 presents the absolute values of the temporal ACFs for the 3D reference model, 3D simulation model with $N_1=N_2=N_3=40$, and 3D simulation result for both low VTD and high VTD scenarios. The temporal ACFs of the 2D simulation model by setting $\beta_{T}^{(n1)} = \beta_{R}^{(n2)} = \beta_{R}^{(n3)} = 0^{\circ}$ are also plotted in Fig.~6. It is clear that no matter what the VTD is, the ACFs of the 2D model always show higher correlation than that of the 3D model. This means that the 2D model overestimates the temporal ACFs. From Fig.~6, we observe that both the 3D simulation model and 3D simulation result closely match the 3D reference model. Moreover, the VTD significantly affects the temporal ACF. In low VTD scenario, the temporal ACF is always higher than that in high VTD scenario. 

\begin{figure} [t]%
\centering
\parbox{8.5cm}{%
\centering \includegraphics[scale=0.46]{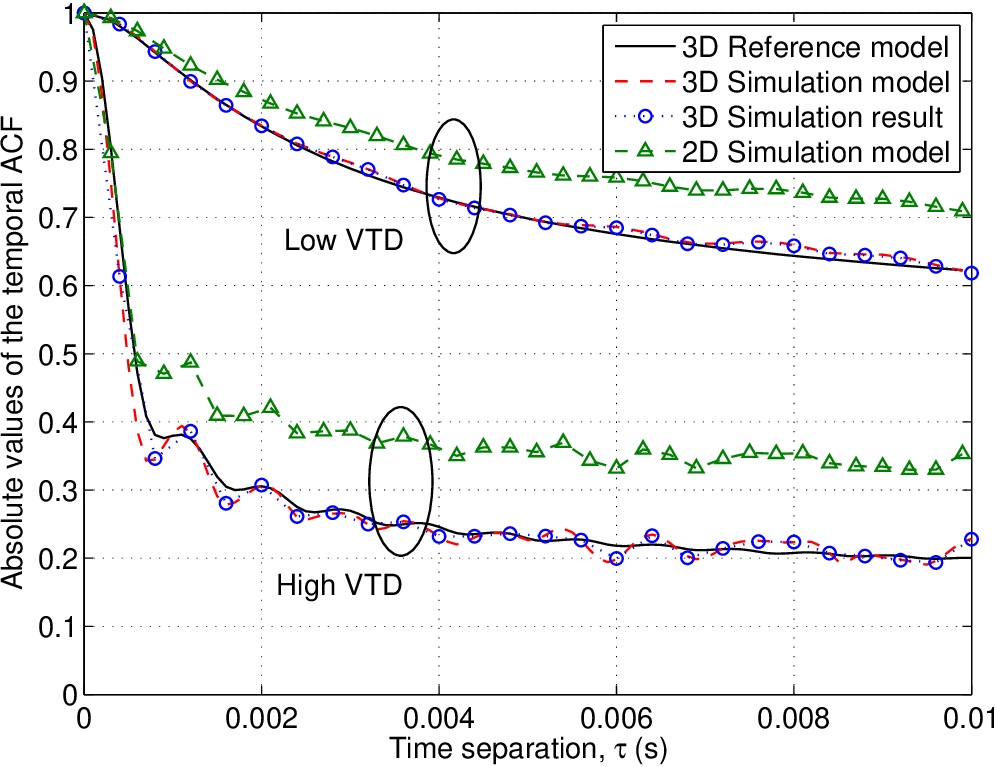} 
\vspace*{-0.2cm} 
\caption{The absolute values of the temporal ACFs for the 3D reference model, 3D simulation model, 3D simulation result, and 2D simulation model ($\delta_T = \delta_R=0$, 3D model: $\beta_{T0}^{(1)}= 6.7^{\circ}$, $\beta_{R0}^{(2)} = 17.2^{\circ}$, $\beta_{R0}^{(3)} = 31.6^{\circ}$, 2D model: $\beta_{T}^{(n1)}$ = $\beta_{R}^{(n2)}$ = $\beta_{R}^{(n3)}$ = $\beta_{T0}^{(1)}$ = $\beta_{R0}^{(2)}$ = $\beta_{R0}^{(3)} = 0^{\circ}$).}%
}%
\quad
\parbox{8.5cm}{%
\centering \vspace{0.5cm}\includegraphics[scale=0.46]{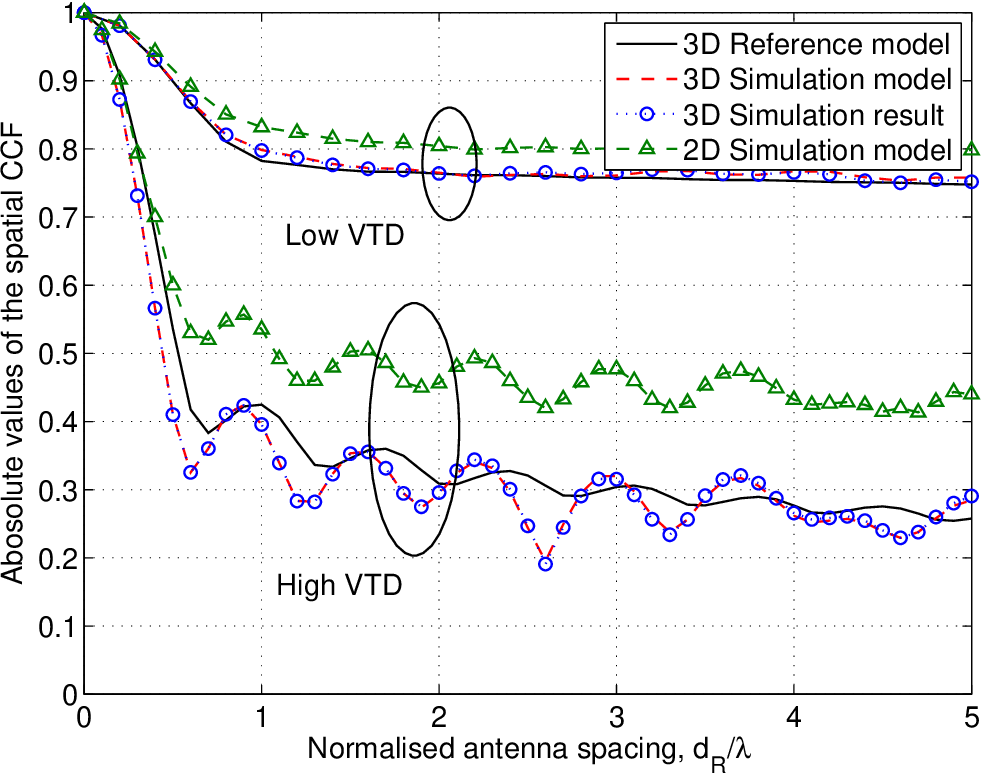} 
\vspace*{-0.2cm}
\caption{The absolute values of the spatial CCFs for the 3D reference model, 3D simulation model, 3D simulation result, and 2D simulation model ($\tau=0$, 3D model: $\beta_{T0}^{(1)} = 6.7^{\circ}$, $\beta_{R0}^{(2)} = 17.2^{\circ}$, $\beta_{R0}^{(3)} = 31.6^{\circ}$, 2D model: $\beta_{T}^{(n1)}$ = $\beta_{R}^{(n2)}$ = $\beta_{R}^{(n3)}$ = $\beta_{T0}^{(1)}$ = $\beta_{R0}^{(2)}$ = $\beta_{R0}^{(3)} = 0^{\circ}$).
}%
}%
\end{figure}%

\subsection{Spatial cross-correlation function}

The spatial cross CF (CCF) can be derived from the ST CF by setting $\tau=0$. Therefore, the spatial CCF can be expressed as 
\begin{align}
\hat{\rho}_{h_{pq}h_{p^\prime q^\prime}}\left(\delta_T,\delta_R \right)=\hat{\rho}_{h_{pq}h_{p^\prime q^\prime}}\left(\delta_T,\delta_R,0\right).
\end{align}

In simulations, the basic parameters are the same as before except for $\delta_T=0.5\lambda$. Fig.~7 presents the absolute values of the spatial CCFs for the 3D reference model, 3D simulation model, 3D simulation result, and 2D simulation model for the low VTD and high VTD scenarios. Both 3D and 2D simulation models have the number of effective scatterers $N_1=N_2=N_3=40$. Again, from Fig.~7, it is clear that higher VTD leads to lower spatial correlation properties. This is because the higher the VTD, the more spatial diversity the V2V channel has. Compared with the 3D models in Fig.~7, 2D simulation model overestimates the spatial correlations. In other words, the 2D model underestimates the spatial diversity gain. The reason is that the 2D model cannot capture the spatial diversity gain in the vertical plane. Moreover, in Figs.~6 and 7 we have shown that 3D simulation results match those of the 3D simulation model very well, indicating the correctness of our derivations. For clarity purposes, we only present 2D and 3D simulation models in the rest of the figures.

\begin{figure} [t]%
\centering
\centering \includegraphics[scale=0.65]{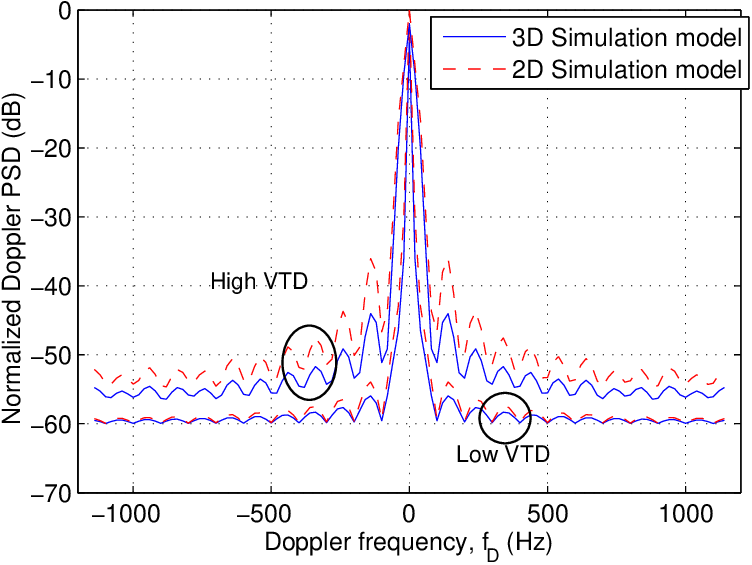} 
\vspace*{-0.3cm} 
\caption{The normalized Doppler PSDs for the 3D and 2D simulation models ($\delta_T=\delta_R=0$, 3D model: $\beta_{T0}^{(1)} = 6.7^{\circ}$, $\beta_{R0}^{(2)} = 17.2^{\circ}$, $\beta_{R0}^{(3)} = 31.6^{\circ}$, 2D model: $\beta_{T}^{(n1)}$ = $\beta_{R}^{(n2)}$ = $\beta_{R}^{(n3)}$ = $\beta_{T0}^{(1)}$ = $\beta_{R0}^{(2)}$ = $\beta_{R0}^{(3)} = 0^{\circ}$). }%
\end{figure}%

\subsection{Doppler PSD}

As the Doppler PSD is derived from the Fourier transform of corresponding temporal ACF, Fig.~8 shows the Doppler PSD of the proposed 3D model compared with 2D one at different VTDs. Comparing the Doppler PSDs with different VTDs in Fig.~8, it shows that the higher the VTD, the more evenly distributed the Doppler PSD is. The underlying physical reason is that in the high VTD scenario, the received power comes from all directions reflected by moving vehicles. However, in the low VTD scenario, the received power comes mainly from specific directions identified by main stationary roadside scatterers and LoS components. Fig.~8 also tells that compared with the 3D model, the 2D model underestimates the Doppler PSD in both low VTD and high VTD scenarios. 

\subsection{Envelope LCR and AFD}

\begin{figure} [t]%
\centering
\parbox{8.5cm}{%
\centering \includegraphics[scale=0.6]{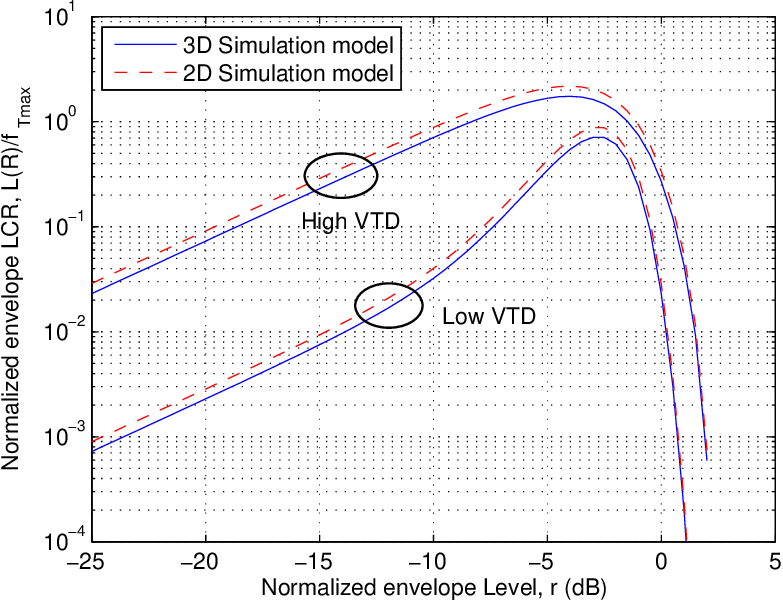} 
\vspace*{-0.1cm} 
\caption{The normalized envelope LCRs for the 3D and 2D simulation models (3D model: $\beta_{T0}^{(1)}$ = $\beta_{R0}^{(2)}$ = $\beta_{R0}^{(3)} = 60^{\circ}$, 2D model: $\beta_{T}^{(n1)}$ = $\beta_{R}^{(n2)}$ = $\beta_{R}^{(n3)}$ = $\beta_{T0}^{(1)}$ = $\beta_{R0}^{(2)}$ = $\beta_{R0}^{(3)} = 0^{\circ}$).}%
}%
\quad
\parbox{8.5cm}{%
\centering \vspace{0.5cm}\includegraphics[scale=0.6]{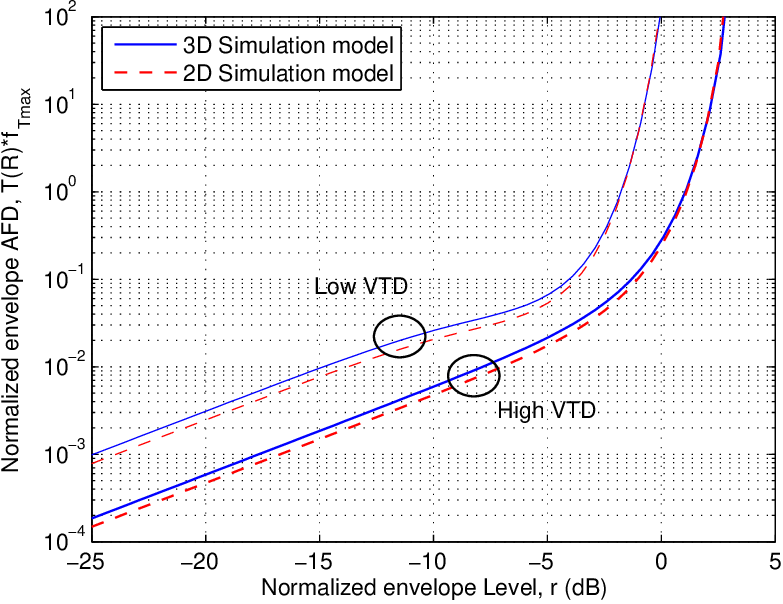} 
\vspace*{-0.1cm}
\caption{The normalized envelope AFDs for the 3D and 2D simulation models (3D model: $\beta_{T0}^{(1)} = \beta_{R0}^{(2)}= \beta_{R0}^{(3)} = 60^{\circ}$, 2D model: $\beta_{T}^{(n1)} = \beta_{R}^{(n2)} = \beta_{R}^{(n3)} = \beta_{T0}^{(1)} = \beta_{R0}^{(2)} = \beta_{R0}^{(3)}=0^{\circ}$).
}%
}%
\end{figure}%

Figs.~9 and 10 depict the envelope LCRs and AFDs for different VTD scenarios (low and high), respectively. Again, the VTD significantly affects the envelope LCR and AFD for V2V channels. Fig.~9 shows that the LCRs are smaller when the VTD is lower. Fig.~10 illustrates that the AFD tends to be larger with lower VTD. However, the elevation angles do not influence the LCR and AFD remarkably. If we used the same elevation parameters (i.e., $\beta_{T0}^{(1)}=6.7^{\circ}$, $\beta_{R0}^{(2)}=17.2^{\circ}$, and $\beta_{R0}^{(3)}=31.6^{\circ}$) as before, the LCR and AFD are barely discernible. The difference is noticeable when we increase the elevation parameters to $\beta_{T0}^{(1)} = \beta_{R0}^{(2)} = \beta_{R0}^{(3)} = 60^{\circ}$ in Figs.~9 and 10. For the envelope LCR in Fig.~9, the 2D model shows higher LCR than the 3D model. For the envelope AFD, the 2D model exhibits smaller AFD than the 3D model. Overall, the elevation angle has minor impact on the envelope LCR and AFD. 

\section{Conclusions} 

In this paper, we have proposed a novel 3D theoretical RS-GBSM and corresponding SoS simulation model for non-isotropic scattering MIMO V2V fading channels. The proposed models have the ability to investigate the impact of the VTD and elevation angle on channel statistics. Furthermore, a novel parameter computation method, named as MEV, has been developed for jointly calculating the azimuth and elevation angles. Based on proposed models, comprehensive statistical properties have been derived and thoroughly investigated. The simulation results have validated the utility of the proposed model. The impact of the elevation angle on channel statistical properties has been investigated and analyzed, i.e., the difference between the 3D and 2D models. By comparing these results, we can see that the VTD has a great impact on all channel statistical properties, whereas the elevation angle has significant impact only on ST CF and Doppler PSD. In addition, our simulations and analysis have clearly addressed that the low VTD condition always shows better channel performance than the high VTD case. Compared with the existing less complex 2D RS-GBSMs and 3D RS-GBSMs, the proposed 3D MIMO V2V RS-GBSMs are more practical to mimic a real V2V communication environment. Our research work can be considered as a theoretical guidance for establishing more purposeful V2V measurement campaigns in the future.

\begin{IEEEbiography}[{\includegraphics[width=1in,height=1.5in,clip,keepaspectratio]{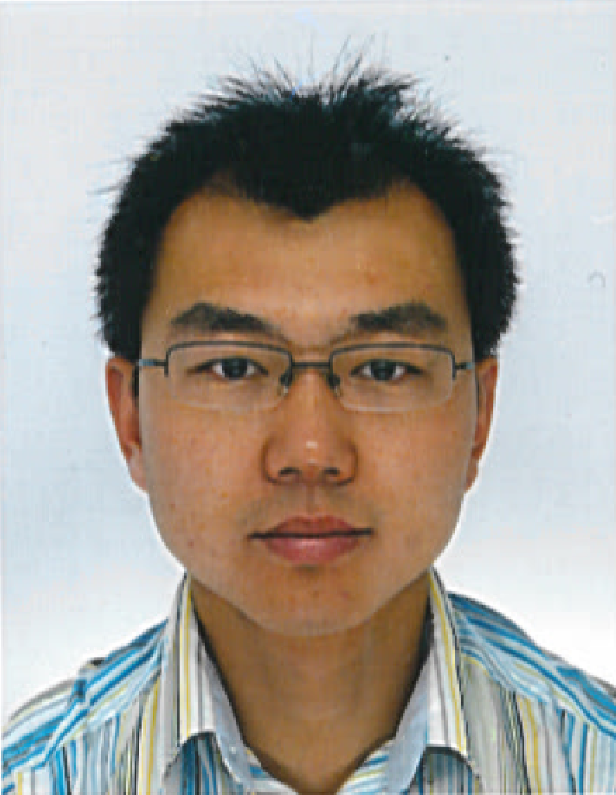}}]
{Yi Yuan} received the BSc degree with distinction in Electronic Engineering from Tianjin University of Technology, Tianjin, China, in 2006 and MSc degree with distinction in Mobile Communications from Heriot-Watt University, Edinburgh, U.K., in 2009. 

He has been a joint PhD student at Heriot-Watt University and The University of Edinburgh since 2010. His main research interests include mobile-to-mobile communications and advanced wireless MIMO channel modeling and simulation.
\end{IEEEbiography}

\begin{IEEEbiography}[{\includegraphics[width=1.5in,height=1.3in,clip,keepaspectratio]{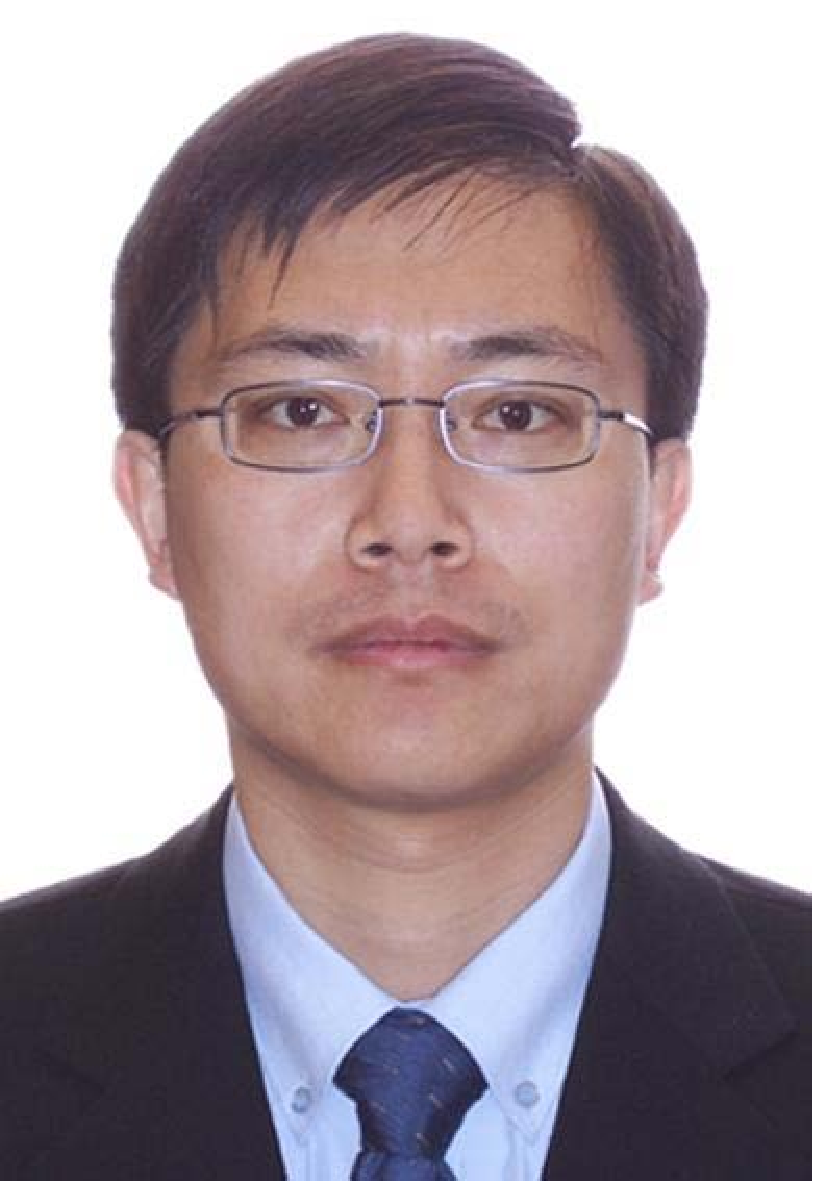}}]
{Cheng-Xiang Wang} (S'01-M'05-SM'08) received the BSc and MEng degrees in Communication and Information Systems from Shandong University, China, in 1997 and 2000, respectively, and the PhD degree in Wireless Communications from Aalborg University, Denmark, in 2004.

He has been with Heriot-Watt University, Edinburgh, U.K., since 2005 and was promoted to a Professor in 2011. He is also an Honorary Fellow of the University of Edinburgh, U.K., and a Chair/Guest Professor of Shandong University and Southeast University, China. He was a Research Fellow at the University of Agder, Grimstad, Norway, from 2001--2005, a Visiting Researcher at Siemens AG-Mobile Phones, Munich, Germany, in 2004, and a Research Assistant at Technical University of Hamburg-Harburg, Hamburg, Germany, from 2000--2001. His current research interests include wireless channel modeling and simulation, green communications, cognitive radio networks, vehicular communication networks, Large MIMO, cooperative MIMO, and 5G wireless communications. He has edited 1 book and published 1 book chapter and over 180 papers in refereed journals and conference proceedings.

Prof. Wang served or is currently serving as an editor for 8 international journals, including IEEE Transactions on Vehicular Technology (2011--) and IEEE Transactions on Wireless Communications (2007--2009). He was the leading Guest Editor for IEEE Journal on Selected Areas in Communications, Special Issue on Vehicular Communications and Networks. He served or is serving as a TPC member, TPC Chair, and General Chair for over 70 international conferences. He received the Best Paper Awards from IEEE Globecom 2010, IEEE ICCT 2011, ITST 2012, and IEEE VTC 2013-Fall. He is a Fellow of the IET, a Fellow of the HEA, and a member of EPSRC Peer Review College.
\end{IEEEbiography}

\begin{IEEEbiography}[{\includegraphics[width=1.5in,height=1.3in,clip,keepaspectratio]{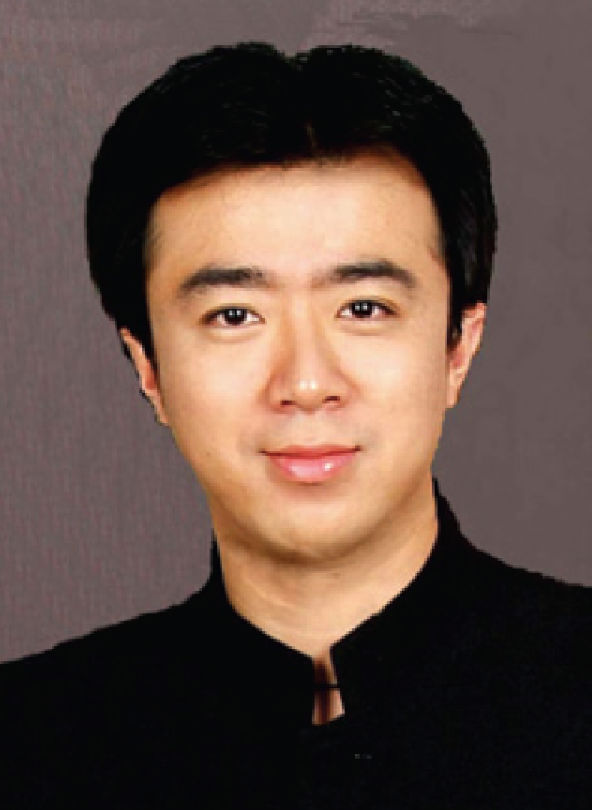}}]
{Xiang Cheng} (S'05-M'10-SM'13) received the PhD degree from Heriot-Watt University and the University of Edinburgh, Edinburgh, U.K., in 2009, where he received the Postgraduate Research Thesis Prize. 

He has been with Peking University, Bejing, China, since 2010, first as a Lecturer, and then as an Associate Professor since 2012. His current research interests include mobile propagation channel modeling and simulation, next generation mobile cellular systems, intelligent transport systems, and hardware prototype development and practical experiments. 

He has published more than 80 research papers in journals and conference proceedings. He received the Best Paper Award from the IEEE International Conference on ITS Telecommunications (ITST 2012) and the IEEE International Conference on Communications in China (ICCC 2013). Dr. Cheng received the ''2009 Chinese National Award for Outstanding Overseas PhD Student" for his academic excellence and outstanding performance. He has served as Symposium Co-Chair and a Member of the Technical Program Committee for several international conferences.
\end{IEEEbiography}

\begin{IEEEbiography}[\vspace{0.0cm}\hspace{-0.0cm}{\includegraphics[width=1.5in,height=1.3in,clip,keepaspectratio]{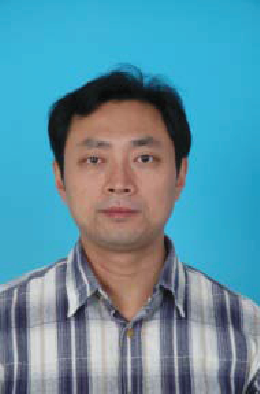}}]
{Bo Ai} (M'01-SM'09) is now working in Beijing Jiaotong University as a professor and PhD advisor. He is vice director of State Key Lab. of Rail Traffic Control and Safety. He is also a vice director of Modern Communications Research Institute. He has authored 5 books and published over 140 scientific research papers in his research area. He has hold 10 national invention patents and 1 US patent. He has been the research team leader for 16 national projects and has won some scientific research prizes such as the First Grade of Technology Advancement Award of Shaanxi Province. 

He is serving as an associate editor for IEEE Transactions on Consumer Electronics and an editorial member of Wireless Personal Communications. He has been notified by Council of Canadian Academies (CCA) that, based on Scopus database, Prof. Ai Bo has been listed as one of the Top 1\% authors in his field all over the world. Prof. Ai Bo has also been Feature Interviewed by IET Electronics Letters. His research interests include radio wave propagation and wireless channel modeling, power amplifier predistortion, LTE-R system and Cyber-physical System. He is an IET Fellow.
\end{IEEEbiography}

\vspace{-11cm}
\begin{IEEEbiography}[\vspace{0.0cm}\hspace{-0.0cm}{\includegraphics[width=1.5in,height=1.3in,clip,keepaspectratio]{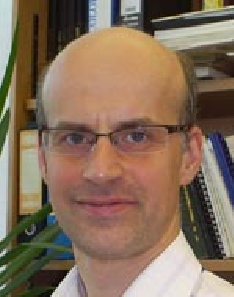}}]
{David I. Laurenson} (M'90) is currently a Senior Lecturer at The University of Edinburgh, Scotland. His interests lie in mobile communications: at the link layer this includes measurements, analysis and modeling of channels, whilst at the network layer this includes provision of mobility management and Quality of Service support. His research extends to practical implementation of wireless networks to other research fields, such as prediction of fire spread using wireless sensor networks, deployment of communication networks for distributed control of power distribution networks, and sensor networks for environmental monitoring and structural analysis.

He is the UK representative for URSI Commission C, and is an associate editor for the Journal of Electrical and Computer Engineering, and the EURASIP Journal on Wireless Communications. He is a member of the IEEE and the IET.
\end{IEEEbiography}

\end{document}